\documentclass[12pt]{article}

\usepackage[USenglish]{babel} 
\usepackage[T1]{fontenc}
\usepackage[ansinew]{inputenc}
\usepackage{verbatim}

\usepackage{lmodern} 
\usepackage{color}

\usepackage{graphicx} 

\usepackage{amsmath}
\usepackage{amsthm}
\usepackage{amsfonts}

\usepackage{hyperref}

\def\a{\alpha}
\def\b{\beta}
\def\c{\gamma}
\def\d{\delta}
\def\e{\epsilon}           
\def\g{\gamma}

\def\k{\kappa}                    
\def\l{\lambda}
\def\m{\mu}
\def\n{\nu}
  
\def\p{\pi}                
\def\r{\rho}                                     
\def\s{\sigma}                                   
\def\t{\tau}

\def\D{\Delta}

\def\G{\Gamma}

\def\L{\Lambda}
\def\O{\Omega}  
\def\P{\Pi}

\def\S{\Sigma}

\def\del{\partial}              
\let\a=\alpha \let\b=\beta \let\g=\gamma \let\d=\delta \let\e=\epsilon
    \let\k=\kappa
\let\l=\lambda \let\m=\mu \let\n=\nu  \let\r=\rho
\let\s=\sigma \let\t=\tau   \let\c=\chi 
  \let\D=\Delta  \let\L=\Lambda
    
\let\ep=\epsilon
    \let\G=\Gamma

\def\nn{\nonumber} \def\bd{\begin{document}} \def\ed{\end{document}}
\def\ds{\documentstyle} \let\fr=\frac \let\bl=\bigl \let\br=\bigr
\let\Br=\Bigr \let\Bl=\Bigl
\let\bm=\bibitem
\let\na=\nabla
\let\pa=\partial \let\ov=\overline
\newcommand{\be}{\begin{equation}}
\newcommand{\ee}{\end{equation}}
\def\ba{\begin{array}}
\def\ea{\end{array}}
\def\ft#1#2{{\textstyle{{\scriptstyle #1}\over {\scriptstyle #2}}}}
\def\fft#1#2{{#1 \over #2}}
\def\del{\partial}
\def\sst#1{{\scriptscriptstyle #1}}
 \def\oneone{\rlap 1\mkern4mu{\rm l}}
\def\ie{{\it i.e.\ }}
\def\via{{\it via}}
\def\semi{{\ltimes}}
\def\str{{\rm str}}
\def\tr{{\rm tr}}
\def\Dm{{{D_{\sst{max}}}}}
\def\vac{ \left | 0 \right \rangle }
\def\kvac{ \left | k \right \rangle }

\def\sp{\; \; \;}

\def\bol{ \left | B (p^+) \right \rangle}
\def\bo1{ \left | B^0 (p^+) \right \rangle}

\def\bolt{ \left | B (p^+) \right \rangle_{\t}}

\def\boxl{ \left | B (x^-) \right \rangle}
\newcommand{\bea}{\begin{eqnarray}}
\newcommand{\eea}{\end{eqnarray}}

\def\<{ \langle }
\def\>{ \rangle }

\def\S{\Sigma}
\renewcommand{\floatpagefraction}{0.6}
\renewcommand{\textfraction}{0.2}

\newcommand\ca{\mathcal{A}}
\newcommand\vp{\varphi}
\newcommand\beal{\begin{align}}
\newcommand\bbone{\ensuremath{\mathbbm{1}}}

\newcommand{\eq}[1]{\begin{equation}#1\end{equation}}
\newcommand{\spl}[1]{\begin{split}#1\end{split}}
\newcommand{\al}[1]{\begin{align}#1\end{align}}
\newcommand{\subeq}[1]{\begin{subequations}#1\end{subequations}}

\newcommand{\arXividhepth}[1]{\href{http://arxiv.org/abs/#1}arXiv:{\tt #1} [hep-th]}
\newcommand{\arXividother}[2]{\href{http://arxiv.org/abs/#1}arXiv:{\tt #1} [#2]}

\newcommand{\bg}[1]{\hat{#1}}
\newcommand{\wj}{\widetilde{J}}
\newcommand{\reo}{\mathrm{Re}~\!\omega}
\newcommand{\imo}{\mathrm{Im}~\!\omega}
\newcommand{\ads}{AdS_4}
\newcommand{\mcal}{\mathcal{M}}
\newcommand{\ccal}{\mathcal{C}}
\newcommand{\ncal}{\mathcal{N}}
\newcommand{\boxedeq}[1]{
\begin{equation}
\fbox{
\rule[0.7cm]{0pt}{0pt}
$#1$
\rule[-0.45cm]{0pt}{0pt}
}
\end{equation}
}

\def\d{\text{d}}

\def\slashchar#1{\setbox0=\hbox{$#1$}           
\dimen0=\wd0                                 
\setbox1=\hbox{/} \dimen1=\wd1               
\ifdim\dimen0>\dimen1                        
\rlap{\hbox to \dimen0{\hfil/\hfil}}      
#1                                        
\else                                        
\rlap{\hbox to \dimen1{\hfil$#1$\hfil}}   
/                                         
\fi}

\def\Re           {{\rm Re\hskip0.1em}}
\def\Im           {{\rm Im\hskip0.1em}}

\newcommand{\E}{\text{\tiny E}}

\begin{document}

\begin{titlepage}

\begin{center}


\vskip 2cm
{\Large \bf Lifshitz as a deformation of Anti-de Sitter}


\vskip 1.25 cm {\bf Yegor Korovin$^1$, Kostas Skenderis$^{1,2,3}$ and Marika Taylor$^{2,3}$}
\\ {\vskip 0.5cm \it\small
KdV Institute for Mathematics$^1$,
Institute for Theoretical Physics$^2$, \\
Science Park 904, 1090 GL Amsterdam, the Netherlands.}
{\vskip 0.2cm \it \small School of Mathematical Sciences and STAG Research Centre$^3$, University of Southampton, Southampton SO17 1BJ, UK.}

{\vskip 0.2cm \small {\it E-mail: J.Korovins@uva.nl, K.Skenderis@soton.ac.uk and  M.M.Taylor@soton.ac.uk} }

\end{center}

\vskip 1 cm

\begin{abstract}
\baselineskip=16pt
We consider holography for Lifshitz spacetimes with dynamical exponent $z=1+\e^2$, where $\epsilon $ is small. We show that the holographically dual field theory is a specific deformation of the relativistic CFT, corresponding to the $z=1$ theory.  Treating $\e$ as a small expansion parameter we set up the holographic dictionary for Einstein-Proca models up to order $\e^2$ in three and four bulk dimensions. We explain how renormalization turns the relativistic conformal invariance into non-relativistic Lifshitz invariance with dynamical exponent $z=1+\e^2$. We compute the two-point function of the conserved spin two current for the dual two-dimensional field theory and verify that it is Lifshitz invariant.
Using only QFT arguments, we show that a particular class of deformations of CFTs generically leads to Lifshitz scaling invariance and we construct  examples of such deformations.
\end{abstract}

\end{titlepage}

\setcounter{tocdepth}{2}

\tableofcontents
\pagebreak








\section{Introduction and summary of results}

There has been considerable interest in recent years in using gauge/gravity
duality as a tool for modelling strongly coupled physics, in particular
with a view to possible applications to condensed matter physics (see \cite{Sachdev:2010ch,McGreevy:2009xe,Horowitz:2010gk,Herzog:2009xv,Hartnoll:2009sz} for reviews). Several interesting condensed matter systems exhibit strongly interacting non-relativistic scale invariant fixed points and
one may hope to use gauge/gravity duality to study them. With
such applications in mind, gravity solutions with
Schr\"{o}dinger \cite{Balasubramanian:2008dm,Son:2008ye}
and Lifshitz \cite{Kachru:2008yh,Taylor:2008tg}
isometries were constructed and studied.

A key idea in holographic approaches
is that the gravity models may allow us to uncover new
universality classes, not easily accessible with perturbation theory,
which may be of relevance to condensed matter physics.
One should note however that
there is very little {\it a priori} evidence that the holographic models
actually describe physics appropriate for the condensed matter systems under question.
The predominant approach has been to proceed phenomenologically
and probe the relevance of these
models by computing observables holographically and comparing to
experimental results. The goal in this paper is rather to
understand better the dual theory from first principles.

When the gravity solution
contains a parameter which can be adjusted such that the solution becomes
a deformation of AdS one may use the standard AdS/CFT correspondence in order
to understand the nature of the dual field theory. This approach was exploited
in \cite{Guica:2010sw} (see also \cite{Son:2008ye,Costa:2010cn,Kraus:2011pf,Guica:2011ia,
vanRees:2012cw})
where it was shown that the theory dual to Schr\"{o}dinger geometries
is a deformation of a relativistic CFT by specific operators which, although
irrelevant from the perspective of the original relativistic CFT,
are exactly marginal from the perspective of the non-relativistic
Schr\"{o}dinger symmetry. Apart from elucidating the holographic
duality for the Schr\"{o}dinger geometries, the analysis in \cite{Guica:2010sw}
also indicated that a new general class of theories with Schr\"{o}dinger
symmetry can be obtained from deformations of relativistic conformal theories, see
\cite{Hofman:2011zj} for a recent application of these ideas.
These deformations do not have to be realized holographically and indeed these results
are interesting for field theory in its own right. In this paper we
will present an analogous set of results for Lifshitz geometries.

There are three different ways to adjust the dynamical exponent such that the theory
becomes a deformation of AdS. Firstly, one can consider the case in which $z$
approaches infinity, when the geometry about which
one is deforming becomes $AdS_2 \times R^{d-1}$. This limit is however not very useful
for our purposes because the holographic dictionary for the limiting spacetime is
not fully understood (due to the non-compact $R^{d-1}$ directions
and the well-known subtleties associated with $AdS_2$ holography). The second case is that of
$z=2$ Lifshitz which can be obtained by a reduction from $z=0$ Schr\"odinger in one dimension
higher \cite{Balasubramanian:2010uk,Costa:2010cn,Narayan:2011az}. These higher dimensional solutions
are asymptotically AdS and therefore holography for $z=2$ Lifshitz can be derived by dimensional reduction,
following \cite{Kanitscheider:2009as,Gouteraux:2011qh}.
This procedure for obtaining the holographic dictionary was followed in  \cite{Chemissany:2011mb,Chemissany:2012du}, reducing the results obtained in \cite{Papadimitriou:2011qb}.
This system has the additional
advantage that it can be embedded in supergravity \cite{Donos:2010tu,Halmagyi:2011xh,Cassani:2011sv,Petrini:2012bh}.
However, the reduction circle becomes null at infinity which implies the dual theory should be
related to the Discretized Light Cone Quantization (DLCQ) of the deformed CFT corresponding to the $z=0$ Schr\"odinger solution, and thus
this approach suffers from the well-known subtleties associated with DLCQ.

The third way to view Lifshitz geometries as being close to AdS is when the dynamical exponent $z$
is near to one. This case is free of any subtleties and will be the topic of this paper.
When the dynamical exponent takes such a value, one can extend the standard
AdS/CFT correspondence in order to understand the dual theory.
Focusing on the pure Lifshitz solution\footnote{In this paper we will focus on exactly Lifshitz solutions but
one could also explore Lifshitz solutions with running scalar couplings (hyperscaling violation), which can for example
be realized using Einstein-Maxwell dilaton models.} of the system of a massive vector
coupled to gravity \cite{Taylor:2008tg}, we find that we can
achieve $z \sim 1 + \e^2$, with $\e$ small, by tuning the mass of the vector field.
Expanding in $\e$ first, the solution indeed becomes asymptotically AdS and
its interpretation can then be extracted from its asymptotics. We find that
the solution is dual to a QFT which is a deformation of a CFT by the time
component of a vector primary operator ${\cal J}^i$ of dimension $d$
(recall that conserved vectors have dimension $\D=(d-1)$),
\be \label{action_def}
S_{Lif} = S_{CFT} + \e \int d^dx {\cal J}^t.
\ee
This is our first key observation. In the deformed theory the vector operator acquires
an anomalous dimension at order $\e^2$ when $d >2$ and at order $\e^4$ when $d=2$.

Note that the most commonly quoted examples of Lifshitz invariant theories are not
of this form. For example, the scalar theory with action
\be
S = \int dt d^3x (\dot{\phi}^2 + \phi (-\partial^2)^z \phi)
\ee
which is often used in the literature as an illustrative
example (especially when $z=2$) does not become of the form (\ref{action_def}) when $z \sim 1 +\e^2$. This suggests
that this field theory model is unlikely to share key features of the holographic model.

A number of theoretical models with dynamical exponents close to one have appeared in the condensed matter literature.
A sample of such models include quantum spin systems with quenched disorder \cite{PhysRevB.26.154}, models for quantum Hall systems
\cite{PhysRevB.50.7526, PhysRevLett.80.5409}, graphene \cite{2006PhRvL..97n6401H, PhysRevB.75.235423}, spin liquids in the presence of non-magnetic disorder \cite{PhysRevB.70.140405} and quantum transitions to and from the superconducting state in high $T_c$ superconductors
\cite{PhysRevLett.85.1532, PhysRevB.61.14723, 2001cond.mat..4053S}.

It turns out that none of these models are governed by an action of the form
(\ref{action_def}) (although some can be viewed as a deformation of a (free) CFT, albeit a different type of deformation). One may then wonder
whether the Lifshitz invariant theories of the form (\ref{action_def}) arise only in holographic
theories at strong coupling or not. In fact we show that the opposite
is true: any deformation of the type (\ref{action}) leads to a
Lifshitz invariant theory with dynamical exponent $z \sim 1+\e^2$,
to leading order in $\e$. This is our second main result.

Let us briefly explain this result: at the classical level the deformation in (\ref{action_def}) breaks
Lorentz invariance but the theory is still invariant under relativistic
scale transformations. Since $\e$ is small one may study the
theory via conformal perturbation theory. The leading order correction
arises at order $\e^2$ and can be computed using the OPE of the vector primary
operator. This OPE contains the following universal terms
\be \label{OPE}
\mathcal{J}_{i}(x) \mathcal{J}_j(0) \sim \frac{I_{ij}}{x^{2 d}} + \cdots + {\cal A}_{ij}{}^{kl} \frac{{\cal T}_{kl}}{x^d} + \cdots,
\ee
where $I_{ij}=\delta_{ij}-2 x_i x_j/x^2$, ${\cal A}_{ij}{}^{kl}$ is some universal tensorial structure and ${\cal T}_{ij}$ is the CFT stress energy tensor.
This follows from the fact that $\mathcal{J}_i$ is a primary operator of dimension $d$ (see section \ref{sec:QFT} for
the detailed argument). The leading order term gives rise to power law divergences
while the term with ${\cal T}_{ij}$ leads to a logarithmic divergence. These divergences
are exactly the same as the ones we find in holographically renormalizing the
gravitational theory. Removing the divergences requires in particular renormalization of
the source of ${\cal T}_{ij}$, i.e. the boundary metric. It turns out that the renormalized metric to this order
is actually equal to the bulk metric at order $\e^2$ with the
cut-off replaced by the radial coordinate! The renormalization leads to a  beta function
for the boundary metric and this modifies the trace Ward identity
which now becomes,
\be \label{Litr}
z \langle  {\cal T}^t_t  \rangle_{\e^2} + \langle {\cal T}^a_a \rangle_{\e^2} =0
\ee
where $i=\{t, a\}, a=1,\ldots,d$ and $\langle\ \rangle_{\e^2}$ denote the
computation up to order $\e^2$. This is precisely the
condition required for Lifshitz invariance!

This construction provides a new, general class of Lifshitz invariant models.
The emergence of Lifshitz invariance is derived using conformal perturbation theory
and relies on a universal part of the OPE of a primary operator of dimension $d$.
The argument is thus valid for any CFT, weakly or strongly coupled. The dimension of
$\mathcal{J}_i$ in the deformed theory however in general depends on the specific details of the CFT. In the holographic
models the dimension of $\mathcal{J}_i$ is not corrected at leading order, which implies that
the 3-point function of $\mathcal{J}_i$ should be zero. Moreover, if the OPE contains other singular terms
beyond the ones exhibited in (\ref{OPE}), these induce relevant deformations of the critical
point and the r.h.s. of (\ref{Litr}) contains corresponding terms.

The holographic model is based on a strongly coupled CFT, but weakly coupled models
may also have interesting applications.
It would be interesting to investigate whether such models could be of relevance
for modelling real world systems (irrespectively of their holographic realization).
As noted above systems with $z$ close to one have already been studied
with a view to applications that range from quantum Hall system to graphene to
high $T_c$ superconductivity, etc. We also note that at long distances Hor\v{a}va-Lifshitz gravity \cite{Horava:2009uw} has a dynamical exponent that approaches $z=1$. We further
add that experimental evidence for critical
behavior with dynamical exponent close to one has been reported in
\cite{PhysRevLett.98.057002, PhysRevLett.95.137002, PhysRevLett.99.237003} for the
transition from the insulator to superconductor in the underdoped region
of certain high $T_c$ superconductors
and in \cite{PhysRevB.83.140507} for the superconductor to metal transition in the overdoped region.

From the bulk side, top down embeddings of Lifshitz solutions with $z \ge 1$ were found in \cite{Gregory:2010gx} and were further explored in \cite{Braviner:2011kz} and \cite{Barclay:2012he}. We will discuss in the conclusions the implications of our results for such top down solutions with $z$ close to one.

This paper is organized as follows. In the next section we present the gravitational solution
with Lifshitz scaling and show that when $z \sim 1+ \e^2$ this model is dual to a deformation
of a CFT by a vector operator of dimension $d$. In section \ref{sec:HoloRen} we develop the holographic
dictionary for our theory. In section \ref{sec:TTcf} the $2$-point function of stress-energy tensor in $2$d Lifshitz invariant theory is computed and in section \ref{sec:QFT} we present the QFT analysis for general dimension. We conclude
in section \ref{sec:con}.

\section{Lifshitz solutions} \label{sec:sol}

Solutions with Lifshitz isometries were first presented in
\cite{Kachru:2008yh}. Here we will use the formulation in
\cite{Taylor:2008tg} in terms of gravity coupled to a massive
vector. The action is given by\footnote{Curvature conventions:
${R_{\m\n\rho}}^{\sigma} = \partial_{\nu} \Gamma^{\sigma}_{\mu \rho} + \G^{\lambda}_{\mu \rho} \G_{\nu \lambda}^{\sigma} - (\mu \leftrightarrow \nu), \
R_{\mu \nu} = {R_{\m\sigma\n}}^{\sigma}$.
}
\be \label{grv_model}
S = \frac{1}{16 \p G_{d+1}} \int d^{d+1}x \sqrt{-G} \left [R + d(d-1) - \frac{1}{4} F^{2} - \frac{1}{2} M^2 A^2 \right ],
\ee
where relative to \cite{Taylor:2008tg} we have rescaled the fields
and the coordinates as,
\be
G^{{\rm here}}_{\mu \nu}
= l^2 G^{{\rm there}}_{\mu \nu}, \qquad A^{{\rm here}}_{\mu}
= l A^{{\rm there}}_{\mu}, \qquad x^\mu_{\rm here} = l x^\mu_{\rm there}
\ee
where
\be
l^2 =\frac{z^2 + z (d-2) + (d-1)^2}{d(d-1)}.
\ee
(We have also absorbed an overall factor of $l^{2d}$ in Newton's
constant.)

The action (\ref{grv_model}) admits an AdS solution with unit AdS radius.
When the mass is equal to
\be \label{mass_rel}
M^2 = \frac{z d (d-1)^2}{z^2 + z (d-2) + (d-1)^2}
\ee
the field equations
also admit a solution with Lifshitz scaling symmetry given by
\bea \label{lif}
ds^2 &=&
dr^2 - e^{2 z r/l} dt^2 + e^{2 r/l} dx^{a} dx_a; \\
A &=& {\cal A} e^{z r/l} dt, \qquad  {\cal A}^2 = \frac{2(z-1)}{z}, \nn
\eea
where $a = 1, \cdots, (d-1)$.
The Lifshitz symmetry is realized by the following transformation,
\be
t \to e^{z\l} t, \qquad x^a \to e^\l x^a, \qquad r \to r - \l l.
\ee

By the standard AdS/CFT dictionary the massive vector (at the AdS
critical point) is dual to a
vector primary operator $J_i$ of dimension
\begin{align}
 \Delta &= \frac{1}{2}(d + \sqrt{(d-2)^2 + 4 M^2})\\
&= \frac{d}{2} + \sqrt{(\frac{d}{2}-1)^2 + \frac{z d (d-1)^2}{z^2 + z (d-2) +(d-1)^2}} \nn
\end{align}
In other words, the action (\ref{grv_model}) governs the dynamics of a
CFT whose spectrum contains a vector primary of dimension $\Delta$.
The same theory contains a Lifshitz critical point if (\ref{mass_rel}),
viewed as an equation for $z$, has real solutions with $z>1$.
In order (\ref{mass_rel}) to have real solutions
the mass has to satisfy,
\be \label{bound}
0 \leq M^2 \leq \frac{d (d-1)^2}{3 d-4}, \qquad (d \geq 2).
\ee
The lower bound comes from unitarity: $M^2 \geq 0$ is equivalent
to the unitarity
bound for vector operators, $\Delta \ge (d-1)$
(the massless case corresponds to a conserved current with $\D=d-1$).
When the bound holds, there are two possible solutions,
\be
z_\pm = 1-\frac{d}{2} +\frac{d (d-1)^2}{2 M^2}
\pm \frac{1}{2 M^2} \sqrt{d (M^2 +(d-1)^2)(d(d-1)^2 - (3d-4)M^2)}
\ee
When $M^2 \sim 0$, $z$ either goes to infinity or to zero, while
when $M^2 \sim d (d-1)^2/(3d-4)$, it approaches $(d-1)$.

It remains to impose the condition $z>1$. One can easily show that
the $z_+ \geq 1$, while $z_-$ starts from 0
at $M^2=0$ and monotonically grows to $z_-=(d-1)$ as $M^2 \to d(d-1)^2/(3 d-4)$.
It follows that it is greater than one only in the range
\be \label{rest_bound}
\frac{(d-1)^2 (8 -3 d + 4 \sqrt{3 d^2 -6 d +4})}{13 d-16} <  M^2 \leq
\frac{d (d-1)^2}{3 d-4},
\ee
where $z_-=1$ at the lower limit and $z_-=(d-1)$ at the upper limit.
We summarize the bounds in Table 1 for up to $d=4$.
\begin{table}
    \begin{tabular}{|l|c|c|}
        \hline
        \ & $z_+>1$ & $z_->1$ \\
        \hline
        $d=2$ & $0<M^2 \leq 1$ & $M^2=1$ \\ \hline
        $d=3$ & $ 0 <M^2 \leq 12/5 \approx 2.4$ & $4/23(-1 + 4 \sqrt{13}) \approx 2.33 <M^2 \leq 12/5 =2.4$
            \\ \hline
        $d=4$ & $0 < M^2 \leq 9/2$ & $-1+2 \sqrt{7} \approx 4.29 <M^2 \leq 9/2$ \\ \hline
    \end{tabular}
\caption{Allowed range of masses for Lifshitz solutions with dynamical exponent $z_+$ and $z_-$ bigger then one. }
\end{table}

It follows from this analysis that a necessary condition for obtaining
a top-down model admitting a Lifshitz realization of the type discussed
here is that the spectrum of the compactification contains massive
vectors with mass in the range (\ref{bound}). If in addition the mass is
within (\ref{rest_bound}) then there are two possible Lifshitz critical
points. These conditions are not sufficient however because they only
guarantee that the quadratic action equals (\ref{grv_model}). In general there would be interaction terms between the massive vector and other modes.
If these interactions are quadratic or higher order in the new field, then one can consistently truncate them. However, if the interactions are linear in the
new fields then the massive vector would source them and they cannot be
ignored. Indeed, in all known consistent truncations that involve massive vectors one needs to
keep additional scalar fields, see for example the first such truncation \cite{Maldacena:2008wh}. This is related to the fact that the
OPE of the vector operator with itself may contain operators other than the stress energy tensor and we will revisit this issue in section \ref{sec:QFT}.

Looking at the spectrum of sphere compactifications one finds that there are
indeed massive vectors in the required range. The spectrum of $AdS_3 \times S^3 \times K3$
\cite{Deger:1998nm} contains a massive vector with $M^2=1$ which leads to $z=1$.  The spectrum of $AdS_4 \times S^7$ \cite{Biran:1983iy} contains
two massive vectors in the allowed range. The first has mass $M^2 =3/4$
and the associated dynamical exponent is $z=(45+3 \sqrt{209})/6 \approx 14.72$
while the second has mass $M^2=2$ and can be associated either with $z=1$
and $z=4$. The spectrum of $AdS_5 \times S^5$ \cite{Kim:1985ez}
contains one vector in allowed range, $M^2=3$, which is associated with
$z=1$ and $z=9$ dynamical exponents.  The cases with $z=1$ do not directly lead to a non-trivial Lifshitz geometry but the dual operator may be used to deform the theory to a Lifshitz point as we describe in section \ref{sec:QFT}. It would be interesting to find whether any such case can be associated with a
consistent truncation.

We will instead here focus on the case where the dynamical exponent is close to one, $z \sim 1 + \e^2$, with $\e \ll 1$. This can be achieved if the mass is
\be
M^2 \sim (d-1) + (d-2) \e^2 + \frac{1+d-d^2}{d(d-1)} \e^4 + \cdots
\ee
which implies that the dual operator has dimension
\be
\Delta = d + \frac{d-2}{d} \e^2 + \frac{(-2 d^3 + 6 d^2 - 7 d + 4)}{d^3 (d-1)}\e^4+\cdots.
\label{anomdim}
\ee
When $M^2=d-1$ the leading order solution is AdS.
Recall that the asymptotic expansion of the bulk vector field is given by
\begin{align}
 A_i = e^{(\Delta-d+1) r} A_{(0)i}+ \cdots + e^{-(\Delta -1) r} A_{(d)i} + \cdots,
\label{expan}
\end{align}
where $A_{(0)i}$ is the source of the dual operator and $ A_{(d)i}$ is related to its expectation value.

We would like now to interpret holographically the Lifshitz solution (\ref{lif}) with $z \sim 1+\e^2$ as a perturbation of AdS with $\e \ll 1$. The metric is $AdS_{d+1}$ up to order $\e^2$ while the massive vector becomes
\begin{align}
A_{(0)t} = \sqrt{2} \epsilon (1+ O(\e^2)).
 \label{lifg}
\end{align}
Thus to order $\e$  the Lifshitz solution has the holographic interpretation as a deformation of the CFT by a vector operator $J_t$ of dimension $d$
\begin{align}
 S_{CFT} \rightarrow S_{CFT} + \sqrt{2} \int{d^dx \epsilon J^t}.
\end{align}
In the following sections we will set up the holographic dictionary for this
case and analyze the dual QFT.

\section{Holographic dictionary} \label{sec:HoloRen}

In this section we will set up the holographic dictionary between the bulk gravity and the dual field theory, working
perturbatively in the
parameter $\ep$. Our aim is to derive the holographic one point functions (in the presence of sources) and the Ward
identities they satisfy.

Holographic renormalization for Lifshitz solutions
was studied in \cite{Ross:2009ar,Ross:2011gu,Baggio:2011cp,Griffin:2011xs,Mann:2011hg,Baggio:2011ha}. In particular, it was
established in \cite{Ross:2011gu}, using the
radial Hamiltonian formalism \cite{Papadimitriou:2004ap,Papadimitriou:2004rz},
that these models can be holographically renormalized for any $z$.
Since the models are non-relativistic it is natural to use the vielbein formalism \cite{Ross:2009ar,Guica:2010sw} and
this is indeed what was used in \cite{Ross:2011gu}.
Here however we will proceed by using the methodology in \cite{deHaro:2000xn}
and the metric formulation. The reason for using the metric formulation is that we are studying the theory from the
perspective of the AdS critical point and the formulation in \cite{deHaro:2000xn} gives the asymptotic
form of the metric in a more direct manner than the radial Hamiltonian
formalism (which in turn is more efficient in producing the counterterms etc.). In contrast to previous approaches we have in mind deforming (by an irrelevant operator) AdS space into Lifshitz and do not assume particular fall-off behaviour for the bulk fields, but derive bulk solution for arbitrary Dirichlet data (see below).

We begin with the action \eqref{grv_model} together with the Gibbons-Hawking boundary term
\begin{align}
 S_{\text{bare}} &= \frac{1}{16 \pi G_{d+1} } \int{d^{d+1} x \sqrt{-G} \Big(R + d(d-1) - \frac{1}{4} F_{\mu \nu} F^{\mu \nu}
- \frac{1}{2} M^2 A_{\mu} A^{\mu} \Big)} \nn
\\
&+ \frac{1}{8 \p G_{d+1}} \int d^{d}x \sqrt{-\g}K,
\label{action}
\end{align}
with $M^2$ given in (\ref{mass_rel}). The associated field equations are
\begin{align}
& D_{\mu} F^{\mu \nu} = M^2 A^{\nu}, \label{Eins} \\
&R_{\mu \nu} = -d G_{\mu \nu} + \frac{M^2}{2} A_{\mu} A_{\nu} + \frac{1}{2} G^{\rho \sigma} F_{\mu \rho} F_{\nu \sigma} +
\frac{1}{4(1-d)} F^{\sigma \lambda} F_{\sigma \lambda} G_{\mu \nu}. \nn
\end{align}
Taking the trace of the Einstein equations and plugging back into \eqref{action} the onshell action is
\begin{align}
\label{onshell}
S_{\text{onshell}} &= \frac{1}{16 \pi G_{d+1}} \int{d^{d+1} x \sqrt{-G} (-2d - \frac{1}{2(d-1)} F_{\mu \nu} F^{\mu \nu})} \\
& +\frac{1}{8 \pi G_{d+1}} \int{d^d x \sqrt{-\g} K}.\nn  \end{align}
This action diverges due to the infinite volume of spacetime and we need to add covariant counterterms in order to render it finite.
The counterterms at order $\e^0$ are well known. Here we would like to compute the required counterterms through order $\e^2$.
To obtain those we need to compute the most general infinities that appear to this order and for this we need to know
the asymptotic solutions of the field equations.

\bigskip

We parameterize the metric and vector field as
\begin{align}
\label{param}
&ds^2 = dr^2 + e^{2r} g_{ij} dx^i dx^j, \\
&g_{ij}(x,r;\e) = g_{[0]ij}(x,r) + \epsilon^2 g_{[2]ij}(x,r) + \ldots \nn \\
& A_i(x,r;\e) = \epsilon e^r A_{(0)i}(x) + \ldots. \nn
\end{align}
For the metric, the notation $g_{[a]ij}$ captures the order in $\e$. Each of these
coefficients has a radial expansion as well and the order in radial expansion
will be denoted (as usual) by curved parentheses. For example,
\be
g_{[0]ij}(x,r) = g_{[0](0)ij}(x) + e^{-2 r} g_{[0](2)ij}(x) + \cdots
\ee
is the radial expansion of the metric at leading order in $\e$. We would like to obtain the most general asymptotic solution given $g_{[0](0)ij}(x)$ and
$A_{(0)i}(x)$ as arbitrary Dirichlet data.

Using this form for the metric the Einstein equations can be expressed as
\begin{align}
\label{Einst}
&Ric[g]_{ij} - e^{2 r} \Big \{ \frac{1}{2} g'' + \frac{d}{2} g' - \frac{1}{2} g' g^{-1} g' + \frac{1}{2} \tr (g^{-1} g')g + \frac{1}{4} \tr(g^{-1} g') g' \Big \}_{ij}  \\ &= \frac{M^2}{2} A_{i} A_{j} +\frac{1}{2}F_{ir}F_{jr}+ \frac{e^{- 2r}}{2} g^{kl} F_{i k} F_{j l} \nn \\&+ \frac{1}{4(1-d)} g^{kl}( 2 F_{kr} F_{lr} + e^{-2r} g^{mn} F_{km} F_{ln}) g_{ij}, \nn\\
& \frac{1}{2} D^j g'_{ij} -\frac{1}{2} D_i \tr(g^{-1} g')=\frac{M^2}{2} A_{i} A_{r} + \frac{e^{- 2r}}{2} g^{jk} F_{i j} F_{r k}, \label{Einstri} \\ 
&-\frac{1}{2} \tr(g^{-1} g'') - \tr(g^{-1} g') + \frac{1}{4} \tr(g^{-1} g' g^{-1} g') \nn \\ &= \frac{M^2}{2} A_{r} A_{r} + \frac{(d-2)}{2(d-1)}e^{- 2r}g^{ij} F_{r i} F_{r j} + \frac{e^{- 4r}}{4(1-d)} F^{i j} F_{ij}\label{Einstrr}
\end{align}
where a prime denotes a derivative with respect to $r$; indices are raised and traces are taken with the metric $g^{ij}$ and
$D_i$ is the covariant derivative constructed from the metric $g_{ij}$.

The vector field equations are
\begin{align}
\label{vecr}
&\partial_i(\sqrt{-g} g^{ij} F_{jr}) = M^2 e^{2r} \sqrt{-g} A_r, \\
&\partial_k(\sqrt{-g} F^{ki})  + e^{(4-d)r}\partial_r(e^{(d-2)r} \sqrt{-g} g^{im} F_{rm})
= M^2 e^{2r}\sqrt{-g} A^i. \label{vec}
\end{align}
Taking the divergence of the last equation we obtain
\begin{align}
& e^{-2r} \partial_i(\sqrt{-g} g^{ij} A_j) +e^{-dr} \partial_r(e^{dr} \sqrt{-g} A_r) \label{div} \\
&=e^{-2r} \partial_i(\sqrt{-g} g^{ij} A_j) + d\sqrt{-g}A_r + \partial_r(\sqrt{-g} A_r)=0. \nn
\end{align}

These equations can now be solved order by order in $\ep$: note that the metric expansion involves even powers of $\ep$ whilst
the vector field is expanded in odd powers. The details of the analysis differ
for different dimensions due to singularities in the numerical coefficients
of the expansion. These imply the appearance of logarithms in special dimensions.
We will thus study the two main cases of interest, $d=2$ and $d=3$, separately.
It is clear that the analysis can be straightforwardly extended to any $d$;
the computations however become increasingly more tedious as we increase the
dimension (as in the standard asymptotically locally AdS cases).

\subsection{Analysis for d=2}

\subsubsection{Zeroth order in $\e$}

The analysis at order $\e^0$ (pure gravity) was done in
\cite{Skenderis:1999nb} leading to the metric
\be
g_{[0]ij}= g_{[0](0)ij}(x) + e^{-2r} g_{[0](2)ij} + e^{-4 r} g_{[0](4)ij}.
\ee
Note that in this case the Fefferman-Graham expansion terminates at order $e^{-4 r}$,
although this will not play any role in this paper.
The coefficient $g_{[0](2)}$ is related to the holographic energy
momentum tensor
\bea
\langle T_{ij} \rangle_{[0]} = -\frac{2}{\sqrt{g_{[0](0)}}} \frac{\delta S_{\text{ren}}}{\delta g_{[0](0)}^{ij}},
\eea
where the subscript indicates
that we work at order $\e^0$, as follows \cite{Skenderis:1999nb,deHaro:2000xn}
\bea
g_{[0](2)ij}&=& \frac{1}{2}\left(-g_{[0](0)ij} R \label{g02}
+ (16 \pi G_3) \langle T_{ij} \rangle_{[0]}\right)  \\
&=&
\frac{1}{2}\left(-g_{[0](0)ij} R
+ \frac{24 \pi}{c}  \langle T_{ij} \rangle_{[0]}\right)\nn
\eea
where the holographic energy momemtum tensor satisfies
\cite{Henningson:1998gx,Henningson:1998ey}
\be
\nabla^j \langle T_{ij} \rangle_{[0]} =0, \qquad
\langle T^i_{i} \rangle_{[0]} = \frac{c}{24 \pi} R
\ee
and $c=3/2 G$ is the Brown-Henneaux central charge \cite{Brown:1986nw} (recall
that we set the AdS radius equal to one).

We also record the holographic counterterms:
\begin{equation}
S_{\text{ct}[0]} = - \frac{1}{8 \pi G_{3}} \int{d^2 x \sqrt{-\g} (1 + \frac{1}{2} R[\gamma] r_0}),
\label{S0}
\end{equation}
where $r_0$ is the radial cutoff.

\subsubsection{First order in $\epsilon$}

At this order the massive vector is non-trivial and
the near-boundary expansions has the form
\begin{align}
&A_i =  e^{r} ( {\cal  A}_{(0)i}(x) + e^{-2r} ({\cal A}_{(2)i}(x) +r \tilde{\cal{A}}_{(2)i}(x)) \ldots ), \label{ex} \\
&A_r =  e^{-r}( A_{(0)r}(x) + e^{-2r} (A_{(2)r}(x) +r a_{(2)r}(x)) \ldots), \nn
\end{align}
where we will take ${\cal A}_{(0) i}(x) = \ep A_{(0) i}(x)$ and work perturbatively in $\ep$.
Note that $A_{(0)r}$ does not represent an independent source; using the divergence equation \eqref{div} one obtains
\begin{align}
 A_{(0)r}(x)
=-\frac{\epsilon}{d-1} \nabla_{i} A^i_{(0)}(x),
\end{align}
where we gave this expression for general $d$ for later use and
$\nabla$ denotes the covariant derivative constructed from
the $g_{[0](0)ij}$
(note that $\nabla_i$ differs from $D_i$ in \eqref{Einstri}!).

It will be useful to define ${\cal A}_{(2) i} = \ep A_{(2) i}$
and $\tilde{{\cal A}}_{(2)}^i = \ep a_{(2)}^i$.
Then the vector field expansion coefficients are
\begin{eqnarray}
\label{w}
a_{(2)}^i &=& \Big[ \frac{1}{2}\nabla_k F^{ki}_{(0)} - \frac{1}{2}\nabla^i(\nabla_j A^j_{(0)})
+ \Big(
\frac{12 \pi }{c}
\langle T^{ij} \rangle_{[0]}
- \frac{R}{4} g_{[0](0)}^{ij}\Big) A_{(0)j} \Big]\\
&=& \Big[\frac{1}{2}\Box A^i_{(0)}- \nabla^i(\nabla_j A^j_{(0)})
+\frac{12 \pi}{c} \langle T^{ij} \rangle_{[0]}
A_{(0)j} - \frac{R}{2} g_{[0](0)}^{ij} A_{(0)j} \Big],\nn \\
a_{(2)r} &=& \ep \nabla_i a_{(2)}^i, \label{q} \\
A_{(2)r} &=& \ep
\Big [-\frac{R}{4} (\nabla_i A_{(0)}^i) + \nabla_i A_{(2)}^i - \frac{1}{2} \nabla^i \nabla_i \nabla_j A^j_{(0)}, \\
&&  + \Big(
\frac{24 \pi}{c}
\langle T^{ij} \rangle_{[0]}
- \frac{R}{2} g_{(0)}^{ij}\Big)\nabla_i A_{(0)j} \Big]. \nn
\end{eqnarray}
Here and later, when we present asymptotic solutions,
 indices are raised using the metric
$g_{[0](0)}^{ij}$.

The coefficient $A_{(2) i}$ is left undetermined by the asymptotic analysis.
As we will see later it is related to the expectation value of the dual
operator, as in the standard case \cite{deHaro:2000xn}. Note further that
the coefficients at order $\e$ depend locally on the sources $g_{[0](0)},
A_{(0)i}$, as expected,. In addition, the coefficients depend also locally on the
zeroth order expectation value of the dual stress
energy tensor $\langle T^{ij} \rangle_{[0]}$, which at first sight appears  problematic
since this coefficient is in general
non-locally related to  $g_{[0](0)}$ and this could lead to non-local divergences. We will see later how this issue is
resolved and the conceptual interpretation of it.


\subsubsection{Second order in $\epsilon$}

Let us consider the backreaction at order $\epsilon^2$.
The asymptotic expansion takes the form
\begin{align} \label{expg[2]}
g_{[2]} = r h_{[2](0)} +re^{-2r} h_{[2](2)}+ e^{-2r} g_{[2](2)}
+ \mathcal{O}(e^{-4r}).
\end{align}
One could also include a term at order $r^0$ in $g_{[2] ij}$,
namely $g_{[2](0) ij}$, but such a  term
would correspond to a modification of the stress energy tensor source
and here we analyze the response of the theory with the  sources kept fixed --
hence this term is set to zero.

We provide details of the computation in Appendix \ref{backr2d}. Here we collect results. $h_{[2](0)ij}$ and $h_{[2](2)ij}$ are determined by the sources (for later use this expression for $h_{[2](0)ij}$ is given for general $d$):
\begin{align}
\label{h20}
h_{[2](0)ij} &= - A_{(0)i}A_{(0)j} + \frac{1}{2 (d-1)}A_{(0)k} A^k_{(0)} g_{[0](0)ij}, \\
\label{h22}
h_{[2](2)ij}
&= \frac{6 \pi}{c}
(A_{(0)k}A_{(0)}^{k} \langle T_{ij} \rangle_{[0]}
- A_{(0)}^{k} \langle T_{kl} \rangle_{[0]} A_{(0)}^{l}g_{[0](0)ij})
+ A_{(0)k}a_{(2)}^{k} g_{[0](0)ij} \nn\\ -&\nabla^k (A_{(0)i}F_{(0)jk}+A_{(0)j}F_{(0)ik}) - \frac{R}{8} A_{(0)k}A_{(0)}^{k} g_{[0](0)ij}+\frac{R}{4} A_{(0)i} A_{(0)j} \\
-& \frac{1}{2}(A_{(0)i} a_{(2)j}+A_{(0)j} a_{(2)i})+ \frac{1}{4} \nabla_k \Big( A^k_{(0)}(\nabla_l A^l_{(0)}) -3A_{(0)l} F^{kl}_{(0)} \Big)g_{[0](0)ij}.\nn
\end{align}
Moreover, divergence of $g_{[2](2)}$ is also fixed by \eqref{diver} and:
\begin{align}
\tr(g_{[2](2)}) \label{trg2}
= A_{(0)i} A^i_{(2)} - \frac{1}{4} A_{(0)i} \Box A^i_{(0)}
-\frac{6 \pi}{c}
A^i_{(0)} \langle T_{ij} \rangle_{[0]} A^j_{(0)}
+ \frac{R}{4}A_{(0)i} A^i_{(0)}.
\end{align}
The part of  $g_{[2](2)ij}$ undetermined by the asymptotic analysis is linked
to the expectaction value of $T_{ij}$ at order $\e^2$, as we will soon find.


\subsubsection{On-shell action and counterterms}

Having computed the most general asymptotic solution through order $\e^2$
we now move on to computing the order $\epsilon^2$ terms in the regulated
on-shell action to which the zeroth order counterterms $S_{\text{ct}[0]}$
have been added.
Computing the divergent terms at order $\ep^2$  we obtain
\bea
&&-\frac{\epsilon^2}{16 \pi G_{3}} \int d^2 x \sqrt{-g_{[0](0)}} \Big[ \frac{e^{2r_0}}{2} A_{(0)i} A^i_{(0)} \label{divergence} \\
&& \qquad + r_0\Big( \frac{12 \pi}{c}
A^i_{(0)}\langle T_{ij} \rangle_{[0]} A^j_{(0)} - \frac{R}{4} A_{(0)i} A^i_{(0)}   \Big) \Big]. \nn
\eea
At first sight this looks problematic since the divergences appear to be
non-local, due to the term involving $\langle T_{ij}\rangle_{[0]}$.
However, this is only a ``pseudo non-local'' divergence, similar to those
discussed in \cite{vanRees:2011fr,vanRees:2011ir}. The non-locality
disappears when we express everything in terms of induced fields at the
regulated surface $r=r_0$.

To find the counterterm action we invert equation \eqref{ex}:
\begin{align}
\e A_{(0)i} &= e^{-r} A_i + \cdots.
\end{align}
Thus the counterterm needed to cancel the leading
order divergence in \eqref{divergence} at order $\epsilon^2$ is
\begin{align}
\tilde{S}_{\text{ct}[2](2)}
&= \frac{1}{32 \pi G_{3}} \int d^2 x \sqrt{-\g} \g^{ij} A_i A_j. \label{lct}
\end{align}
Noting that
\begin{align}
\sqrt{-\g} \g^{ij} A_i A_j=\e^2 \sqrt{-g_{[0](0)}} (g^{ij}_{[0](0)}A_{(0)i}A_{(0)j} e^{2r} + 2rA_{(0)i} a^i_{(2)} )+\ldots
\end{align}
we see that this counterterm cancels the apparently non-local piece
in \eqref{divergence} involving $\langle T_{ij}\rangle_{[0]}$.

After canceling the leading order divergence we are left with a logarithmic
divergence (which originates from the leading order counterterm):
\begin{align}
&\frac{\epsilon^2}{16 \pi G_{3}} \int d^2 x \sqrt{-g_{[0](0)}}  r_0\Big( \frac{1}{2}(\nabla_i A^i_{(0)})^2 - \frac{1}{4} F_{(0)ij}F^{ij}_{(0)} \Big),
\end{align}
which in turn can be cancelled by the logarithmic counterterm
\begin{align}
\tilde{S}_{\text{ct}[2](0)}= -\frac{1}{16 \pi G_{3}} \int d^2 x \sqrt{-\g} r_0 \Big( \frac{1}{2}(\nabla_i A^i)^2 - \frac{1}{4} F_{ij}F^{ij} \Big).
\label{logct}
\end{align}
Thus, in summary, the total counterterm action in two dimensions becomes
\begin{align}
S_{\text{ct}}=S_{\text{ct}[0]}+ S_{\text{ct}[2]} &= - \frac{1}{8 \pi G_{3}} \int{d^2 x \sqrt{-\g} (1 + \frac{1}{2} R r_0) } \\
&+\frac{1}{32 \pi G_{3}} \int d^2 x \sqrt{-\g} \g^{ij} A_i A_j \nn \\&-\frac{1}{16 \pi G_{3}} \int d^2 x \sqrt{-\g} r_0 \Big( \frac{1}{2}(\nabla_i A^i)^2
- \frac{1}{4} F_{ij}F^{ij} \Big). \nn
\end{align}
These counterterms suffice to render the action finite to order $\ep^2$.

\subsubsection{Computation of renormalized one-point functions}

To compute the renormalized one point functions we need to vary the total action.
Note that the variation of $\delta g^{ij}$ includes variation of the vector source $\delta A_{(0)j}$:
\begin{align}
\delta g^{ij}
&=\delta g_{[0](0)}^{ij} \label{delg} \\
&+\e^2 r \Bigg(2 g_{[0](0)}^{jk} A^i_{(0)} \delta A_{(0)k} - \delta A_{(0)k}A^k_{(0)} g_{[0](0)}^{ij}- \frac{1}{2} A_{(0)k}A^k_{(0)} \delta g_{[0](0)}^{ij}\nn \\ & - \frac{1}{2} A_{(0)k}A_{(0)l} g_{[0](0)}^{ij} \delta g_{[0](0)}^{kl}+ A^i_{(0)}A_{(0)k}\delta g_{[0](0)}^{kj} + A^j_{(0)}A_{(0)k}\delta g_{[0](0)}^{ki} \Bigg) \nn
\end{align}
The variation of the renormalized action with respect to the vector source is finite by construction and is given by
\begin{align}
\frac{\ep^2}{16 \pi G_{3}} \int d^2 x \sqrt{-g_{[0](0)}} g^{ij}_{[0](0)} (2 A_{(2)i} - a_{(2)i} - \nabla_i \nabla_j A_{(0)}^j) \delta A_{(0)j} .
\end{align}
The corresponding vector $1$-point function is defined as
\be
\langle {\cal J}^i \rangle =-\frac{1}{\sqrt{-g_{[0](0)}}} \frac{\delta S_{\text{ren}}}{\delta {\cal A}_{(0)i}} =-
 \frac{1}{\sqrt{-g_{[0](0)}}} \frac{\delta S_{\text{ren}}}{\ep \delta A_{(0)i}}
\ee
and is given by $\langle {\cal J}^i \rangle = \ep \langle J^i \rangle$ with
\begin{align}
\langle J^i \rangle =-\frac{1}{16 \pi G_{3}} (2 A_{(2)}^i - a_{(2)}^i - \nabla^i \nabla_j A_{(0)}^j).
\label{J}
\end{align}
As expected the part of the asymptotic expansion, $A_{(2)}^i$, undetermined by asymptotics is directly related with the 1-point function
of the dual operator.

Now let us compute $1$-point function of the stress-energy tensor:
\be
\langle {\cal T}_{ij} \rangle = \langle T_{ij}\rangle_{[0]} + \ep^2 \langle T_{ij}\rangle_{[2]} + \cdots
\ee
The stress-energy tensor $1$-point function to order $\ep^2$ is obtained using the finite part of the action variation at order $\ep^2$:
\begin{align}
&\frac{1}{16 \pi G_{3}} \int d^2 x \sqrt{-g_{[0](0)}} \\&
\quad \delta g_{[0](0)}^{ij}\Big[\frac{1}{2}h_{[2](2)ij} - g_{[2](2)ij} -\Big(\frac{1}{2}\tr(h_{[2](2)} - g_{[0](2)}h_{[2](0)}) - \tr(g_{[2](2)}) \Big)g_{[0](0)ij}  \nn \\
&+A_{(0)i} A_{(2)j}- \frac{1}{2} A_{(0)k} A^k_{(2)}g_{[0](0)ij} + \frac{1}{4}A^k_{(0)} g_{[0](2)kl} A^l_{(0)} g_{[0](0)ij} - \frac{1}{4}A_{(0)k} A^k_{(0)} g_{[0](2)ij}\Big] . \nn
\end{align}

After using \eqref{h22} we obtain the correction to the stress energy tensor at order $\ep^2$:
\begin{align}
\langle T_{ij}\rangle_{[2]} &= -\frac{2}{\sqrt{-g_{[0](0)}}}\frac{\delta S_{[2]\text{ren}}}{\delta g^{ij}_{[0](0)}} \label{T} \\
&=\frac{1}{8 \pi G_3} \Big[ g_{[2](2)ij} -A_{(0)i} A_{(2)j} - \frac{1}{2} A_{(0)k} A^k_{(2)} g_{[0](0)ij}  -\frac{1}{2} (\nabla_k A^k_{(0)})^2 g_{[0](0)ij}\nn\\
&-\frac{R}{8}A_{(0)i} A_{(0)j}+\frac{R}{16}A_{(0)k} A^k_{(0)} g_{[0](0)ij}+\frac{1}{2}A_{(0)k} a^k_{(2)} g_{[0](0)ij} \nn\\
&+\frac{1}{4}(A_{(0)i} a_{(2)j} + A_{(0)j} a_{(2)i}) +\frac{1}{8} \nabla_k \Big(A_{(0)l}F_{(0)}^{kl} + 5 A^k_{(0)} \nabla_l A^l_{(0)} \Big)g_{[0](0)ij}  \nn\\
& - \frac{1}{4}\Big( F_{(0)i}{}^{k} F_{(0)jk} + A_{(0)i} \nabla^k F_{(0)kj} + A_{(0)j} \nabla^k F_{(0)ki}\Big) \Big]. \nn
\end{align}
Again, as expected, the expectation value of ${\cal{T}}_{ij}$ is directly related with the undetermined coefficient, $g_{[2](2)ij}$.

These 1-point functions were obtained using minimal subtraction. We will shortly see that they can be somewhat simplified in a different scheme, where certain
finite boundary terms are added to the action. We will first analyze the Ward identities, however.

\subsubsection{Ward identities} \label{sec:WId2}

The divergence of the order $\e^2$ contribution to stress-energy tensor can be obtained using equation \eqref{diver}:
\begin{align}
\nabla^j \langle T_{ij}\rangle_{[2]} & = \frac{1}{16 \pi G_3} \Big[ 2A_{(2)}^j F_{(0)ij} - 2 A_{(0)i} (\nabla_j A^j_{(2)}) + \frac{12 \pi}{c} \langle T_{jk} \rangle_{[0]} \nabla^k(A_{(0)i} A^j_{(0)}) \nn\\ &-\frac{12 \pi}{c}  A^k_{(0)} \langle T_{kl} \rangle_{[0]} \nabla_i A^l_{(0)}-\frac{1}{4}  A_{(0)i} \nabla_j (R A^j_{(0)}) +\frac{1}{2}A_{(0)i} \nabla^k \nabla_k \nabla_jA^j_{(0)}\nn\\ &- \frac{1}{2} F_{(0)ij}\nabla^k \nabla_k A^j_{(0)}   \Big] \nn\\
&= A_{(0)i} \nabla_j \langle J^j \rangle - \langle J^j \rangle F_{(0)ij}.
\end{align}
The complete energy momentum tensor then satisfies
\be
\nabla^j \langle {\cal T}_{ij}\rangle = {\cal A}_{(0)i} \nabla_j \langle {\cal J}^j \rangle - \langle {\cal J}^j \rangle {\cal F}_{(0)ij}.
\ee
where we recall that  ${\cal A}_{(0)i} = \e  A_{(0)i},    \langle {\cal J}^j \rangle = \ep \langle J^j \rangle$ and
${\cal F}_{(0)ij}$ is the field strength of ${\cal A}_{(0)i} $.
This is precisely the correct diffeomorphism Ward identity.  The terms in the rhs represent the contribution due to the coupling of
vector operator (see for example \cite{Bianchi:2001kw}). There is no explicit $\e$ dependence in this equation and this suggests it may hold beyond the small $\e$ limit.

Let us now turn to the trace identity.
Computing $\langle T^i_{i}\rangle_{[2]}$ leads to
\begin{align}
\label{Ward}
\langle T^i_{i}\rangle_{[2]} &=- \frac{1}{8 \pi G_3} \Big[A_{(0)i} A_{(2)}^i
- \frac{12 \pi}{c} A^i_{(0)}\langle T_{ij} \rangle_{[0]} A^j_{(0)}
+ \frac{1}{4} F_{(0)ij}F_{(0)}^{ij}\\&+ \frac{R}{4}A^i_{(0)} A_{(0)i}-\frac{1}{4}\nabla_k \Big( A_{(0)i}F_{(0)}^{ki} + A^k_{(0)}(\nabla_i A^i_{(0)}) \Big)\Big] \nn\\
&= A_{(0)i} \langle J^i \rangle + \mathcal{A}, \nn
\end{align}
where we used \eqref{J} and
\be
\label{tranomaly}
\mathcal{A} = \frac{1}{16 \pi G_3} \left(\frac{12 \pi}{c}
A^i_{(0)} \langle T_{ij}\rangle_{[0]} A^j_{(0)} - \frac{1}{4} F_{(0)ij}F_{(0)}^{ij} + \frac{1}{2}(\nabla_i A^i_{(0)})^2 - \frac{R}{4}A^i_{(0)} A_{(0)i}\right)
\ee
This is expected form of an anomalous trace Ward identity. The first term
in the rhs of (\ref{Ward}) is due to the fact that we deformed the theory by a dimension
$2$ vector operator. The second term $\mathcal{A}$ is the correction to the
trace anomaly and
it must be a Weyl invariant.

In fact this term is related to the Weyl invariant action given in \cite{Deser:1983mm} which in our conventions and in arbitrary dimension $d>2$
can be written as
\begin{align}
 \mathcal{L} = -\frac{1}{4} F_{ij} F^{ij} - \frac{d-4}{2 d} (\nabla_i A^i)^2 + \frac{d-4}{2} S_{ij} A^i A^j -\frac{d-4}{8 (d-1)} R A_i A^i, \label{L}
\end{align}
where
\be \label{Sch}
S_{ij} = \frac{1}{d-2} \Big(R_{ij} - \frac{R}{2 (d-1)} g_{ij} \Big)
\ee
is the Schouten tensor. This action is only valid for $d>2$ because of the
singularity in (\ref{Sch}) as $d \to 2$. We note however that when $d \geq 3$,
$S_{ij}=-g_{[0](2)}$ (see (A.1) of \cite{deHaro:2000xn}, reproduced here for $d=3$
in \eqref{g2}) and moreover if in $d=2$ we replace $S_{ij}$ by
$-g_{[0](2)ij}$ (given by (\ref{g02})) we get $\mathcal{A}$!

One can check the Weyl-invariance of $\mathcal{A}$ by direct computation.
The Weyl transformation of $\langle T_{ij} \rangle_{[0]}$ is
well known: $\delta \langle T_{ij} \rangle_{[0]} = \frac{c}{12 \pi}(\nabla_i \nabla_j \s - g_{(0)ij} \Box \s)$.
This is a standard result in CFT and it has also been derived holographically in
\cite{deHaro:2000xn}.
With this information at hand one can check that under a Weyl transformation such that $\delta g_{[0](0)ij} = 2 \s g_{[0](0)ij}$ and
$\delta A_{(0)i} = \s A_{(0)i}$
\begin{align}
&\delta \Big( \frac{12 \pi}{c}A^i_{(0)} \langle T_{ij} \rangle_{[0]} A^j_{(0)} - \frac{1}{4} F_{(0)ij}F_{(0)}^{ij} + \frac{1}{2}(\nabla_i A^i_{(0)})^2 - \frac{R}{4}A^i_{(0)} A_{(0)i}\Big) \\&= - 2 \s \Big( \frac{12 \pi}{c}A^i_{(0)} \langle T_{ij} \rangle_{[0]} A^j_{(0)} - \frac{1}{4} F_{(0)ij}F_{(0)}^{ij} + \frac{1}{2}(\nabla_i A^i_{(0)})^2 - \frac{R}{4}A^i_{(0)} A_{(0)i}\Big) \nn
\end{align}
up to a total derivative term. Thus the anomaly term in $\langle T^i_{i}\rangle_{[2]}$ is indeed Weyl invariant.

The complete anomaly through order $\e^2$ is given by
\bea \label{trd2}
&&\langle {\cal T}^i_i \rangle
- \frac{1}{2} {\cal A}_{(0)}^i \langle {\cal T}_{ij} \rangle  {\cal A}_{(0)}^j
\\&& \qquad \quad
= {\cal A}_{(0)i} \langle {\cal J}^i \rangle - \frac{c}{2 4 \pi} \left( -R +
\frac{1}{4} {\cal F}_{(0)ij} {\cal F}_{(0)}^{ij}-\frac{1}{2}(\nabla_i {\cal A}^i_{(0)})^2 + \frac{R}{4}{\cal A}^i_{(0)}{\cal  A}_{(0)i} \right) \nn
\eea
where this formula holds through order $\e^2$. As in the case of the diffeomorpshim Ward identity, the $\ep$ dependence is implicit
and this formula may hold away from the small $\ep$ limit.

As we will discuss in more detail  in section \ref{sec:QFT}, the term quadratic in ${\cal A}_{(0)}^i$ on the lhs can be thought of as a beta function
contribution to the trace Ward identity, where the beta function is that of the
source of ${\cal{T}}_{ij}$, namely of the metric $g_{(0)ij}$.
Indeed the asymptotic expansion of the bulk metric contains a leading order logarithic
term at order $\e^2$, the $h_{[2](0)}$ term in (\ref{expg[2]}), which can be thought of as renormalizing the leading order metric.

\subsubsection{Scheme dependence}

One can simplify the one-point
function of the vector operator by adding finite counterterms.  The details of the computation are given in appendix \ref{ap3};
here we only summarize the results.
Requiring that $\langle J^i \rangle$ contains only non-local terms can be achieved by adding
\begin{align}
 S_{\text{ct, finite}} = -\frac{1}{8}  \int \sqrt{\g} \left( F_{ij} F^{ij} +
2 (\nabla_i A^i)^2 + R A^2\right).
\end{align}
Then
\begin{align}
\langle {\cal J}^i \rangle = -\frac{2 \e}{16 \p G_3} A^i_{(2)} + \frac{1}{2} \langle T^{ij} \rangle_{[0]}  {\cal A}_{(0)j}.
\label{Jimp}
\end{align}
The first term in this formula is in agreement with the result of \cite{Costa:2010cn} (upon continuation $z \rightarrow 1$). The second term is related to the OPE coefficient of two ${\cal J}^i$'s  to ${\cal T}^{kl}$.

The addition of these finite counterterms modifies also the one-point function of the stress-energy tensor, which becomes
\begin{align}
\langle T_{ij}\rangle_{[2]}
&= \frac{1}{2} A^k_{(0)} \langle T_{kl}\rangle_{[0]} A^l_{(0)} g_{[0](0)ij} -
\frac{1}{8 \pi G_3} \Big[ -g_{[2](2)ij} +A_{(0)i} A_{(2)j}  \label{Tmp} \\
&+ \frac{1}{2} A_{(0)k} A^k_{(2)} g_{[0](0)ij}-\frac{1}{4}(A_{(0)i} a_{(2)j} + A_{(0)j} a_{(2)i})+\frac{R}{16}A_{(0)k} A^k_{(0)} g_{[0](0)ij}   \nn\\&-\frac{5}{8} A^k_{(0)} \nabla_k \nabla_l A^l_{(0)} g_{[0](0)ij} - \frac{1}{4} (\nabla_k A^k_{(0)})^2 g_{[0](0)ij}+ \frac{1}{2} A_{(0)j} \nabla_i \nabla_k A^k_{(0)} \nn\\
& + \frac{1}{4}\Big(A_{(0)i} \nabla^k F_{(0)kj} + A_{(0)j} \nabla^k F_{(0)ki}\Big)- \frac{3}{8}A_{(0)l} \nabla_k  F_{(0)}^{kl} \Big].\nn
\end{align}
The addition of finite counterterms does not affect the Ward identities.

\subsubsection{Recovering the Lifshitz invariance} \label{sec:RecLif}

Let us now fix the source terms to be those corresponding to the  Lifshitz solution in  \eqref{lif} with $z=1+\e^2$:
\be \label{Lif_sou}
{\cal A}_{(0)t} = \sqrt{2}\e, \qquad g_{[0](0)ij} = \eta_{ij}.
\ee
The trace Ward identity (\ref{trd2}) becomes
\be \label{Lifin}
z \langle {\cal T}^t_t \rangle +  \langle {\cal T}^x_x \rangle = \e \sqrt{2} \langle {\cal J}^t \rangle,
\ee
where the contribution of the term quadratic in ${\cal A}_{(0)i}$ in the lhs of (\ref{trd2})  led to the change the coefficient of $\langle {\cal T}^t_t \rangle$
from 1 to $z=1+\e^2$ . When $\langle {\cal J}^t \rangle =0$  then  (\ref{Lifin}) is precisely the condition for Lifshitz invariance !
(We will review this in section \ref{sec:QFT}). When $\langle {\cal J}^t \rangle \neq 0$ the Lifshitz invariance is spontaneously
broken.

Let us now evaluate the holographic formulas for the solution  in  \eqref{lif} with $z=1+\e^2$.
Expanding first in $\e$ we find that $ g_{[0](2)} = 0$ which implies $\langle T_{ij}\rangle_{[0]}=0$. Furthermore,
expanding the vector field we find  $A_{(2)i}=0$ and using (\ref{Jimp}) we conclude
\be
 \langle {\cal J}^i \rangle = 0,
\ee
so indeed we recover the Lifshitz symmetry from the QFT Ward identity (\ref{Lifin}).

Expanding the metric in $\e$ also gives $h_{[2](0)tt}= -\e^2 = h_{[2](0)xx}$, $h_{[2](2)}=0=g_{[2](2)}$. Thus $h_{[2](0)}$ is traceless in agreement with our general results (see equation \eqref{h20}).
Using \eqref{Tmp} we finally obtain
\be
\langle {\cal T}_{ij}\rangle = 0,
\ee
which is indeed what we would expect, since this geometry should be dual to the (scale invariant) vacuum of the Lifshitz theory. In \cite{Korovin2}
we will discuss the black hole solution in the deformed theory, which corresponds to a thermal state in the
Lifshitz theory.

\subsection{Analysis for d=3}

\subsubsection{Zeroth order in $\epsilon$}

The analysis at order $\e^0$ was carried out in \cite{deHaro:2000xn} and we summarize the results here.
The asymptotic expansion of the metric  is given by
\be
g_{[0]ij} = g_{[0](0) ij} + e^{-2r} g_{[0](2) ij}+ e^{-3r}g_{[0](3) ij} + \cdots
\ee
Here $g_{[0](2) ij}$ is determined in terms of $g_{[0](0) ij}$
\begin{align}
&g_{[0](2) ij} =  -R_{ij} + \frac{R}{4} g_{[0](0) ij}, \label{g2}
\end{align}
while $g_{[0](3) ij}$ is traceless and divergenceless and is related to the holographic stress energy tensor by
\be \label{3dT0}
\langle T_{ij} \rangle_{[0]} = \frac{3}{16 \pi G_4} g_{[0](3) ij}.
\ee
The gravitational counterterms are given by
\begin{align}
S_{\text{ct}[0]} = -\frac{1}{16 \p G_{4}}\int d^3 x \sqrt{-\g} \Big(4 +R[\g] \Big).
\end{align}

\subsubsection{First order in $\e$}

Only the vector has a contribution linear in $\ep$:
\begin{align}
&A_i = e^{r}({\cal A}_{(0)i}(x) + e^{-2r} {\cal A}_{(2)i}(x) ++ r e^{-3r} \tilde{\cal A}_{(3) i}+ e^{-3r} {\cal A}_{(3)i}(x)
+ \ldots), \\
&A_r = e^{-r}(A_{(0)r}(x) + e^{-2r} A_{(2)r}(x) + r e^{- 3r} a_{(3) r} + e^{-3 r} A_{(3)r}(x) + \ldots).
\end{align}
and (as in the $d=2$ case) we will assume the source is linear in $\e, {\cal A}_{(0)i}(x)=\epsilon A_{(0)i}(x)$.
It is also useful to define  ${\cal A}_{(2)}^i = \ep A_{(2)}^i$ and $\tilde{\cal A}_{(3) i} = \ep a_{(3) i}$.

Solving asymptotically the field equations we find for the spatial components of the vector,
\begin{align}
A^i_{(2)} &= \frac{1}{2(d-2)} \Big( \nabla_k F^{ki}_{(0)} -(d-3)\nabla^i A_{(0)r} - (\tr(g_{[0](2)})A^i_{(0)} - 2 g_{[0](2)}^{ij} A_{(0)j})\Big)\nn\\
&\stackrel{d=3}{=} \frac{1}{2} \Big(\nabla_k F^{ki}_{(0)} + \frac{3 R}{4} A^i_{(0)} - 2 Ric^{ij} A_{(0)j}\Big)\\
a_{(3)i} &= g_{[0](3) ij} A_{(0)}^j, \nn
\label{a3exp}
\end{align}
while the radial components are given by
\begin{align}
A_{(0)r} &= - \frac{\ep}{2}\nabla_i A_{(0)}^i; \qquad
A_{(2)r} = -\frac{\ep}{4}\Big( \Box + \frac{R}{4} \Big)\nabla_i A_{(0)}^i +\nabla_i A_{(2)}^i,\\
a_{(3)r} &= \ep \nabla_i a_{(3)}^i, \qquad
A_{(3)r} = \ep \nabla_i A_{(3)}^i.
\end{align}

\subsubsection{Second order in $\e$}

Next let us consider the backreaction on the metric to order $\e^2$. The asymptotic expansion of the $\ep^2$ term in the metric is
\begin{align}
g_{[2]ij} = r h_{[2](0)ij} + r e^{-2r} h_{[2](2)ij} +e^{-2r} g_{[2](2)ij} + r e^{-3r}h_{[2](3)ij} + e^{-3r}g_{[2](3)ij} + \ldots
\end{align}
where we set to zero the possible contribution to the source at order $\e^2$.
Using the expansions of the field equations given in appendix \ref{Eq-analysis}, one can express these coefficients as follows. The leading
term $h_{[2](0)ij}$ is given by \eqref{h20}.
The terms at order $r$ in \eqref{Rij} give us the expression for $h_{[2](2)ij}$:
\begin{align}
 h_{[2](2)ij} &= \frac{1}{4} \Big[(\nabla_k A_{(0)l})(\nabla^k A_{(0)}^l) - (\nabla_k A_{(0)}^k)^2 + 2 A_{(0)k} \Box A_{(0)}^{k} - 2 A_{(0)}^{k} \nabla_k \nabla_l A_{(0)}^{l}  \nn \\
&+ \frac{1}{2} F_{(0)kl} F_{(0)}^{kl} \Big] g_{[0](0)ij}  -\frac{3}{8} \nabla_i \nabla_j (A_{(0)k} A_{(0)}^k)- \frac{R}{4}A_{(0)i}  A_{(0)j}  \\&+ \frac{1}{2} \nabla^k \Big(\nabla_i(A_{(0)j}A_{(0)k}) + \nabla_j(A_{(0)i}A_{(0)k}) - \nabla_k(A_{(0)i}A_{(0)j})  \Big). \nn
\end{align}
From the terms at order one in \eqref{Rij}:
\begin{align}
 g_{[2](2)ij} &= - \frac{R}{8} A_{(0)i}  A_{(0)j} + A_{(0)i} R_{jk} A_{(0)}^k - \frac{1}{4} A_{(0)k} A_{(0)}^k Ric_{ij}+\frac{3}{16} \nabla_i \nabla_j (A_{(0)k} A_{(0)}^k) \nn\\&- \frac{1}{4}\nabla^k \Big(\nabla_i(A_{(0)j}A_{(0)k}) + \nabla_j(A_{(0)i}A_{(0)k}) - \nabla_k(A_{(0)i}A_{(0)j})  \Big) +\frac{1}{2} F_{(0)i}{}^{k} F_{(0)jk} \nn\\
&+\frac{1}{4} \Big( A_{(0)i} \nabla_j \nabla_k A_{(0)}^k + A_{(0)j} \nabla_i \nabla_k A_{(0)}^k + A_{(0)i} \nabla^k F_{(0)kj} + A_{(0)j} \nabla^k F_{(0)ki} \Big) \nn \\
&+\Big[ \frac{3}{16}(\nabla_k A_{(0)}^k)^2 - \frac{1}{4} F_{(0)kl} F_{(0)}^{kl} - \frac{3}{8} A_{(0)k} \Box A_{(0)}^k - \frac{1}{4} A_{(0)}^k \nabla_l \nabla_k  A_{(0)}^l \\&-\frac{1}{8}(\nabla_k A_{(0)l}) (\nabla^k A_{(0)}^l) +\frac{1}{2} A_{(0)}^k \nabla_k \nabla_l  A_{(0)}^l +\frac{3}{16}R A_{(0)k} A_{(0)}^k \Big]g_{[0](0)ij}. \nn
\end{align}
Using $e^{-r}$ terms in \eqref{Rij} we find
\begin{align}
h_{[2](3)ij} &= \frac{3}{8} A_{(0)k} A_{(0)}^k g_{[0](3)ij} - \frac{2}{3} A_{(0)}^k \Big( g_{[0](3)ik}A_{(0)j}+ g_{[0](3)jk} A_{(0)i} \Big) \\&+ \frac{1}{3} A_{(0)}^k g_{[0](3)kl} A_{(0)}^l g_{[0](0)ij}. \nn
\label{h23}
\end{align}
The term $g_{[2](3) ij}$ is left undetermined by the asymptotic analysis, up to trace and divergence constraints. Its trace is
\begin{equation}
 \tr(g_{[2](3)}) = \frac{2}{3} A_{(0)i} A_{(3)}^i - \frac{5}{18} A_{(0)}^i g_{[0](3)ij} A_{(0)}^j,
\end{equation}
whilst its divergence satisfies
\begin{align}
 &\nabla^j h_{[2](3)ij} - 3 \nabla^j g_{[2](3)ij} -g_{[0](3)}^{jk} \nabla_k h_{[2](0)ij} - \frac{1}{2}h_{[2](0)}^{jk} \nabla_i g_{[0](3)jk} \\&- \nabla_i \tr(h_{[2](3)} - 3 g_{[2](3)} - g_{[0](3)} h_{[2](0)}) \nn\\&= 2 (A_{(3)i} A_{(0)r} + A_{(3)r} A_{(0)i}) -g_{[0](3)}^{jk} F_{(0)ij} A_{(0)k} \nn\\&+  A_{(0)}^j (\nabla_i A_{(3)j} - \nabla_j A_{(3)i}) - 2 A^j_{(3)} F_{(0)ij} +  a^j_{(3)} F_{(0)ij}. \nn
\label{dg23}
\end{align}

\subsubsection{Counterterms}

Evaluating the on-shell action we find that the leading order divergence is given by
\begin{align}
S_{\text{div}} = \frac{\ep^2}{16 \pi G_4}  \int d^3 x \sqrt{-g_{[0](0)}} e^{3r} \Big( -\frac{1}{2} A_{(0)i} A_{(0)}^i \Big) + \cdots
\end{align}
There is in addition a logarithmic divergence containing the non-local combination $A_{(0)}^i  g_{[0](3)ij} A_{(0)}^j$:
\begin{align}
S_{\text{div}} = - \frac{\ep^2}{16 \pi G_4}  \int d^3 x \sqrt{-g_{[0](0)}} r \Big( 2 A_{(0)}^i  g_{[0](3)ij} A_{(0)}^j \Big).
\end{align}
Both these divergences are removed by the local counterterm
\begin{align}
S_{\text{ct} [2]} = \frac{1}{32 \pi G_4} \int d^3 x \sqrt{-\g} A_{i} A^i.
\label{lc}
\end{align}
Thus, we find again that all the counterterms are local.

\subsubsection{Renormalized one-point functions}

Although there are additional local counterterms required to remove subleading divergences at order $\ep^2$, only this leading counterterm \eqref{lc}
can contribute finite pieces to the $1$-point functions. Therefore we do not need to
construct explicitly all the counterterms: it is enough to know that these are local. Defining the one point function of the vector operator as
\be
\langle {\cal J}^{i} \rangle =-\frac{1}{\sqrt{-g_{[0](0)}}} \frac{\delta S_{\text{ren}}}{\delta {\cal A}_{(0) i}}= -\frac{1}{\sqrt{-g_{[0](0)}}} \frac{\delta S_{\text{ren}}}{\ep \delta A_{(0) i}},
\ee
and using
\begin{align}
(16 \p G_4) \delta_A S_{\text{on-shell}} = -\int d^3 x \sqrt{-\g} \g^{ij} F_{ri} \delta A_{j} + \int d^3 x \sqrt{\g} \g^{ij} A_{i} \delta A_{j}
\end{align}
we compute
\begin{align}
\left< {\cal J}^i \right> =-\frac{3 \ep}{16 \p G_4 }  A_{(3)}^i  + \frac{1}{3} \langle T^{ij} \rangle_{[0]} {\cal A}_{(0)j},
\label{3dJ}
\end{align}
where we used (\ref{3dT0}).

From
\begin{align}
16 \p G_4 \delta_{\g} S_{\text{on-shell}} = &\int d^3 x \sqrt{-\g} (K_{ij} - K \g_{ij} + 2 \g_{ij}) \delta \g^{ij} \\+& \frac{1}{2} \int d^3 x \sqrt{-\g} (A_i A_j - \frac{1}{2} A_k A^k \g_{ij}) \delta \g^{ij} \nn
\end{align}
one can compute the stress energy tensor
\be
\left< {\cal T}_{ij} \right> =-\frac{2}{\sqrt{-g_{[0](0)}}} \frac{\delta S_{\text{on-shell}}}{\delta g^{ij}_{[0](0)}}.
\ee
Then
\be
\left< {\cal T}_{ij} \right> = \langle T_{ij} \rangle_{[0]} + \ep^2 \langle T_{ij} \rangle_{[2]} + \cdots
\ee
where the leading order term is given in (\ref{3dT0}) whilst
\begin{align}
\label{3dTimp}
\left< T_{ij} \right>_{[2]}
&= -\frac{1}{16 \p G_4}\Big[  h_{[2](3)ij} - 3 g_{[2](3)ij} -\frac{1}{4} A_{(0)k} A_{(0)}^k g_{[0](3)ij} \\
&-A_{(0)}^k g_{[0](3)kl} A_{(0)}^l g_{[0](0)ij}+ A_{(0)k} A_{(3)}^k g_{[0](0)ij} + A_{(0)i} A_{(3)j} + A_{(0)j} A_{(3)i}  \Big]. \nn
\end{align}

\subsubsection{Ward identities}

Using \eqref{dg23} we can check that the expected diffeomorphism Ward identity is satisfied:
\begin{align}
\nabla^j \left< T_{ij} \right>_{[2]} = A_{(0)i} \nabla_j \left< J^j \right> - \left< J^j \right> F_{(0)ij}.
\end{align}
where we used $\langle {\cal J}^{i} \rangle = \ep \left< J^i \right> $.

For the trace of stress-energy tensor we obtain
\begin{align}
 \left< T_{i}^i \right>_{[2]} &= -\frac{1}{16 \p G_4} (3 A_{(0)i} A_{(3)}^i - \frac{5}{2} A_{(0)}^i g_{[0](3)ij} A_{(0)}^j) \\ &= A_{(0)i} \left< J^i \right> + \frac{1}{2} A_{(0)}^i\langle T_{ij} \rangle_{[0]} A_{(0)}^j. \nn
\label{3dT}
\end{align}
Thus through order $\e^2$ the complete trace Ward identity is
\be \label{trd3}
\langle {\cal T}^i_i \rangle - \frac{1}{2} {\cal A}_{(0)}^i \langle {\cal T}_{ij} \rangle  {\cal A}_{(0)}^j
= {\cal A}_{(0)i} \langle {\cal J}^i \rangle.
\ee
This is precisely the same as the $d=2$ case in (\ref{trd2}), except that in $d=2$ we have additional terms related to the
conformal anomaly.  The terms quadratic in ${\cal A}_{(0)i}$
can be thought of as a beta function contribution to the trace Ward identity.

\subsubsection{Recovering Lifshitz invariance}

The discussion parallels that given for $d=2$ in section \ref{sec:RecLif}. The quadratic terms in $ {\cal A}_{(0)}^j$
are responsible for producing the Lifshitz Ward identity
\be
z \langle {\cal T}^t_t \rangle +  \langle {\cal T}^x_x \rangle = \e \sqrt{2} \langle {\cal J}^t \rangle,
\ee
once we set the sources to the values relevant for pure Lifshitz (\ref{Lif_sou}).

Furthermore, the 1-point functions evaluated on the pure Lifshitz solution  \eqref{lif} with $z=1+\e^2$ yield
\be
 \langle {\cal J}^i \rangle = 0, \qquad \langle {\cal T}_{ij}\rangle = 0.
\ee
Thus the geometry can be interpreted as the vacuum state in the Lifshitz theory, as anticipated.

\section{Lifshitz invariant correlation functions in two dimensions} \label{sec:TTcf}

To derive higher-point correlation functions in general spacetime dimension one needs to solve the bulk field equations around the background. However in $d=2$ the situation simplifies: as we review below, in a relativistic two-dimensional CFT conformal symmetry is powerful enough to fix the $2$-point function of stress-energy tensor with itself. (In fact, all higher point functions of stress-energy tensor are fixed by the symmetry). We show here that a similar argument applies to the Lifshitz invariant theory.

\subsection{Correlation functions in the relativistic CFT}
Let us start with the relativistic theory. Consider metric fluctuations $g_{\m\n} = \eta_{\m\n} + h_{\m\n}$. The diffeomorphism and trace Ward identities in standard complex coordinates are given by
\begin{align}
\label{dW}
\bar{\pa} \left\langle T_{ww}\right\rangle + \pa \left\langle T_{w \bar{w}}\right\rangle &=0,\\
\label{tW}
\left\langle T_{w \bar{w}}\right\rangle &= \frac{1}{4}\cdot \frac{c}{24 \p} R[h]  \nn \\&=\frac{1}{4}\cdot \frac{c}{24 \p} (\pa^2 h^{ww} + \bar{\pa}^2 h^{\bar{w} \bar{w}} -  \bar{\pa} \pa h^{w \bar{w}} -  \bar{\pa} \pa h^{w \bar{w}}),
\end{align}
where in the last equality we have linearised the Ricci scalar which suffices for $2$-point functions.

The main idea of the argument is then the following. By taking further functional derivatives of the diffeomorphism Ward identity we obtain a system of differential equations for different $2$-point correlation functions. For example, differentiating (\ref{dW}) with respect to $h^{ww}$  yields
\begin{equation}
\label{difeq}
\bar{\pa} \left\langle T_{ww}(w) T_{ww}(0)\right\rangle = - \pa \left\langle T_{w \bar{w}} (w) T_{ww}(0)\right\rangle.  
\end{equation}
Then we use the trace Ward identity to compute one of the terms (involving the trace) and integrating these differential equations we obtain the $2$-point functions.

Indeed, taking the functional derivative of \eqref{tW} with respect to $h^{ww}$ we find
\begin{equation}
\left\langle T_{w \bar{w}} (w)  T_{w {w}} (0) \right\rangle = \frac{1}{4}\cdot \frac{c}{24 \p} 4 \pa^2 \delta^2(w,\bar{w})
=\frac{c}{24 \p} \pa^2 \frac{1}{2\p} \bar{\pa} \pa \log |w|^2,
\end{equation}
and using \eqref{difeq} we obtain
\begin{equation}
\bar{\pa} \left\langle T_{ww}(w) T_{ww}(0)\right\rangle = - \pa \left\langle T_{w \bar{w}} (w) T_{ww}(0)\right\rangle \\
= - \frac{c}{24 \p} \frac{1}{2\p} \bar{\pa} \pa^4 \log |w|^2.
\end{equation}
Integrating this equation (which amounts to cancelling the factor of $\bar{\pa}$ in the left and right hand side) we find the well-known result:
\be
\left\langle T_{ww}(w) T_{ww}(0)\right\rangle = \frac{1}{(2 \p)^2} \frac{c/2}{w^4}.
\ee
(The factor $1/(2\p)^2$ follows from our  ``bulk'' convention for the energy momentum tensor, i.e. $T_{\m\n} = \frac{2}{\sqrt{g}} \frac{\delta S}{\delta g^{\m\n}}$, see the discussion around (\ref{defT}).)
A similar computations yields the 2-point function $\left\langle T_{\bar{w}\bar{w}}(\bar{w}) T_{\bar{w}\bar{w}}(0)\right\rangle = 1/(2 \p)^2 (c/2)/\bar{w}^4$
in the anti-holomorphic sector.
Finally, functionally differentiating (\ref{tW}) w.r.t. $h^{\bar{w} w}$ we find
the 2-point function
$\left\langle T_{w \bar{w}}(w,\bar{w})) T_{w \bar{w}}(0)\right\rangle=-\frac{c}{24 \p} \pa \bar{\pa} \delta^2(w,\bar{w})$, which is a contact term as expected.

\subsection{Lifshitz  theory}

Now we move on to the Lifshitz case. Our goal is to obtain the $2$-point function by exploiting the underlying symmetry. Let us decompose $A = A^B + a$, where $A^B$ is the constant background value of order $\e$ necessary to support the Lifshitz geometry and $a$ is a fluctuation around this background. Similarly, we write $g_{\m\n} = \eta_{\m\n} + h_{\m\n}$.

The diffeomorphism Ward identity can be rewritten in complex coordinates ($w$, $\bar{w}$) as
\be
\bar{\pa} \left\langle T_{ww}\right\rangle + \pa \left\langle T_{w \bar{w}}\right\rangle = A^B_{(0)w} (\pa \left\langle J_{\bar{w}}\right\rangle + \bar{\pa} \left\langle J_{w}\right\rangle) + \ldots,
\ee
where we omitted terms which do not contribute to the two-point functions in the scale invariant vacuum (i.e. the vacuum with $a=0$).

It is convenient to define a conserved spin two current by
\be
J_{\m\n} = T_{\m\n} - A^B_{\m} J_{\n}.
\ee
In the scale-invariant vacuum we have
\begin{align}
&\bar{\pa} \left\langle J_{ww}\right\rangle + \pa \left\langle J_{w \bar{w}}\right\rangle =0, \label{vacdif} \\
&\bar{\pa} \left\langle J_{\bar{w}w}\right\rangle + \pa \left\langle J_{\bar{w} \bar{w}}\right\rangle =0, \nn
\end{align}
i.e. this operator is conserved. 

The trace Ward identity reads
\begin{align}
\label{WI}
\left\langle T_{w \bar{w}}\right\rangle = &\frac{1}{2}(A^B_{[0]w} \left\langle J_{\bar{w}}\right\rangle + A^B_{[0]\bar{w}} \left\langle J_{{w}}\right\rangle) \\&+ \frac{1}{4}\frac{c}{24 \p} R[h] + \frac{1}{2} \left\langle T_{ww}\right\rangle (A^B_{[0]\bar{w}})^2 + \frac{1}{2} \left\langle T_{\bar{w} \bar{w}}\right\rangle (A^B_{[0]{w}})^2.  \nn 
\end{align}
Similarly to the  relativistic case we can use \eqref{WI} to compute one of the $2$-point functions.
Differentiating this identity with respect to $a_w$ we get 
\begin{align}
\left\langle T_{w \bar{w}} J_{\bar{w}} \right\rangle = &\frac{1}{2}A^B_{[0]{w}} \left\langle J_{\bar{w}} J_{\bar{w}}\right\rangle + \mathcal{O}(\e^2),  
\label{TJ}
\end{align}
where we have dropped terms which vanish in the scale-invariant vacuum; note also that $\left\langle J_{w} J_{\bar{w}} \right\rangle$ is of order $\e$ or higher. 

Similarly, differentiating \eqref{WI} with respect to $h_{w \bar{w}}$ and using \eqref{TJ} we get up to quadratic order in $A^B_{[0]w}$
\begin{align} \label{TT}
&\left\langle T_{w \bar{w}}T_{w \bar{w}}\right\rangle = \frac{1}{4}\Big((A^B_{[0]w})^2 \left\langle J_{\bar{w}} J_{\bar{w}}\right\rangle + (A^B_{[0]\bar{w}})^2 \left\langle J_{{w}} J_{{w}}\right\rangle  \Big)  + \left\langle T_{w \bar{w}}T_{w \bar{w}}\right\rangle_{CFT} \\
& \qquad  \qquad  \quad+\frac{1}{2} \Big((A^B_{[0]\bar{w}})^2 \left\langle T_{w \bar{w}}T_{w w}\right\rangle + (A^B_{[0]{w}})^2 \left\langle T_{w \bar{w}}T_{\bar{w} \bar{w}}\right\rangle \Big).  \nn
\end{align}
Note that we kept several contact terms, like $\left\langle T_{w \bar{w}}T_{w \bar{w}}\right\rangle_{CFT}$ (see the previous subsection).
Using \eqref{TJ} and \eqref{TT} we finally compute
\begin{align}
\left\langle J_{w \bar{w}} (w) J_{w \bar{w}}(0)\right\rangle &= \left\langle T_{w \bar{w}}(w)T_{w \bar{w}}(0)\right\rangle, \\ 
\left\langle J_{w \bar{w}} (w) J_{\bar{w} w}(0)\right\rangle &= \left\langle T_{w \bar{w}} (w) T_{\bar{w} w } (0)\right\rangle - \frac{1}{2}\Big((A^B_{[0]w})^2 \left\langle J_{\bar{w}} J_{\bar{w}}\right\rangle + (A^B_{[0]\bar{w}})^2 \left\langle J_{{w}} J_{{w}}\right\rangle \Big). \nn
\end{align}
Let us now return to the diffeomorphism Ward identity \eqref{vacdif}. This provides us with a set of differential equations for correlation functions. Consider differentiating it with respect to the source to which $J_{\m\n}$ couples. One gets
\begin{align}
\label{relations}
&\bar{\pa} \left\langle J_{ww}(w) J_{ww}(0)\right\rangle = -  \pa \left\langle J_{w \bar{w}}(w) J_{w {w}}(0)\right\rangle, \\
\label{rrelations}
 &\bar\pa \left\langle J_{ww}(w) J_{w \bar{w}}(0)\right\rangle=-\pa \left\langle J_{w \bar{w}}(w) J_{w \bar{w}}(0)\right\rangle. 
\end{align}
One can obtain similar relations by complex conjugating \eqref{relations} and \eqref{rrelations}, which amounts to making the replacements $\pa \leftrightarrow \bar{\pa}$ and $w \leftrightarrow \bar{w}$.

This information suffices to derive all two-point correlation functions of $J_{\m\n}$. Inserting the background value for vector field\footnote{Our Euclidean conventions are $w = x + i t_E$, $\bar{w} = x - i t_E$ following those of \cite{Polchinski:1998rq}.}: $A_{(0)t} = \sqrt{2} \e$ or $A_{(0)w} =-A_{(0)\bar{w}}= \e/\sqrt{2}$ and using
$\left\langle J_w(w)  J_w(0)\right\rangle = -(C_J/2)/(w^3 \bar{w})$ and $\left\langle J_{\bar{w}} (w) J_{\bar{w}}(0)\right\rangle =$ $ -(C_J/2)/(w \bar{w}^3)$
we directly compute
\begin{align}
\label{jwbwjwbw}
&\left\langle J_{w \bar{w}}(w)J_{w \bar{w}}(0)\right\rangle = -\e^2\frac{C_J }{16} \Big( \frac{1}{w^3 \bar{w}} +\frac{1}{w \bar{w}^3} \Big) + \left\langle T_{w \bar{w}}(w)T_{w \bar{w}}(0)\right\rangle_{CFT}  \\
&\qquad \qquad  \qquad \qquad + \frac{\e^2}{4} \Big(\left\langle T_{w \bar{w}}(w)T_{ww}(0)\right\rangle + \left\langle T_{w \bar{w}}(w)T_{\bar{w} \bar{w}}(0)\right\rangle \Big), \nn\\
\label{jwbwjbww}
&\left\langle J_{w \bar{w}}(w)J_{\bar{w}w }(0)\right\rangle = \left\langle J_{w \bar{w}}(w)J_{w \bar{w}}(0)\right\rangle+\e^2\frac{C_J }{8} \Big( \frac{1}{w^3 \bar{w}} +\frac{1}{w \bar{w}^3} \Big) 
\end{align}
Note that we keep local terms because they are needed for the derivation of subsequent formulae. Using \eqref{rrelations} we can integrate \eqref{jwbwjwbw} once to obtain
\begin{align}
\label{JwwJwbw}
\left\langle J_{ww}(w) J_{w \bar{w}}(0)\right\rangle &= -\e^2\frac{C_J }{16} \Big( -\frac{1}{2 w^2 \bar{w}^2} +3\frac{\log |w|^2}{w^4} \Big) +\frac{\e^2}{4} \frac{1}{(2 \p)^2} \frac{c/2}{w^4} \\&-\frac{\pa}{\bar{\pa}}\left\langle T_{w \bar{w}}(w)T_{w \bar{w}}(0)\right\rangle_{CFT} +\frac{\e^2}{4}\left\langle T_{w w}(w)T_{\bar{w} \bar{w}}(0)\right\rangle + \e^2 F(w) \nn,
\end{align}
where we have explicitly introduced a local term needed for Lorentz invariance (at order $\e^0$) and absorbed other local terms in holomorphic function $F(w)$. 
Integrating \eqref{relations} we obtain
\bea
\label{JwwJww}
\left\langle J_{ww}(w) J_{w w}(0)\right\rangle &=& -\e^2\frac{C_J }{16} \Big( \frac{1}{ w^3 \bar{w}} +12\frac{\bar{w} \log |w|^2- \frac{5}{4}\bar{w}}{w^5} \Big) \\ &+&\e^2 \frac{c/2}{(2 \p)^2} \frac{\bar{w}}{w^5}+\frac{1}{(2\p)^2} \frac{c/2}{w^4} +\e^2(-\bar{w}\pa F(w) + G(w)), \nn
\eea
where $G(w)$ is also local. A completely analogous reasoning applies to the antiholomorphic components.

Likewise, starting from \eqref{jwbwjbww} one can compute $\left\langle J_{ww}(w) J_{\bar{w}w}(0)\right\rangle$ and other correlators. 
Omitting all details we just quote the results: 
\begin{align}
\left\langle J_{w \bar{w}}(w)J_{\bar{w} \bar{w}}(0)\right\rangle &= \left\langle J_{ww}(w) J_{ \bar{w} w}(0)\right\rangle^{\ast}, \\
\left\langle J_{w {w}}(w)J_{\bar{w} \bar{w}}(0)\right\rangle &= \left\langle J_{w \bar{w}}(w)J_{\bar{w}w }(0)\right\rangle,  \\
\left\langle J_{ww}(w) J_{\bar{w} w }(0)\right\rangle &= - \left\langle J_{ww}(w) J_{w \bar{w}}(0)\right\rangle +\frac{\e^2}{2} \frac{c/2}{(2 \p)^2} \frac{1}{w^4}, 
\end{align}
up to local terms which in particular involve the $T_{\mu \nu}$ correlator in the CFT with the same index structure. Here $\ast$ denotes complex conjugation and all other correlators can be obtained
by complex conjugation.

\subsection{Lifshitz invariance of two-point functions}
To obtain the scaling properties of the conserved stress-energy tensor $J_{\m\n}$ under the Lifshitz rescaling 
\be
\label{resc}
x \rightarrow \l x,\qquad t \rightarrow \l^z t
\ee
it is more convenient to rewrite the correlators in Cartesian coordinates, using the results in the previous subsection. For example, we obtain
\begin{align}
 &\left\langle J_{tt}(t,x) J_{tt}(0)\right\rangle = (1-\e^2)2\frac{c/2}{(2 \p)^2} \frac{(t^4 - 6 t^2 x^2 + x^4)}{(t^2 + x^2)^4} -\frac{1}{6} \e^2 \frac{c/2}{(2 \p)^2}\frac{t^2 -x^2}{(t^2 + x^2)^3} \nn \\&+2\e^2 \frac{c/2}{(2 \p)^2} \frac{(t^2 - x^2) (t^4 - 14 t^2 x^2 + x^4)}{(t^2 + x^2)^5} (\log(t^2 + x^2) -\frac{9}{4})+ \ldots, 
\end{align}
where we have omitted scheme dependent (local) terms. To simplify the result we have also explicitly used $c = 6 \pi^2 C_J= 3/(2 G_3)$, where $C_J$ is obtained from \eqref{J2ptfn}. 

On general grounds one expects that $2$-point correlation functions in a Lifshitz invariant theory take the form
\be
 \left\langle \mathcal{O}_{\D_{L1}}  \mathcal{O}_{\D_{L2}} \right\rangle = x^{-\D_{L1} - \D_{L2}} f(\c),
\ee
where the function $f$ depends on the ratio $\c=t/x^{z}$ but is otherwise in general undetermined by the scale invariance. Our results take exactly this form, for example
\begin{align}
\label{ttttcorr}
&\left\langle J_{tt}(t,x) J_{tt}(0)\right\rangle = \frac{c}{(2 \p)^2}  x^{-2(1+z)} \Big[ (1-\e^2)\frac{ \c^4 - 6\c^2 +1 }{(\c^2 +1)^4}\\
&-\frac{1}{12}\e^2 \frac{\chi^2 - 1}{(\chi^2 +1)^3}+\e^2 \frac{(\c^2-1) (\c^4 - 14 \c^2 +1)}{(\c^2+1)^5} (\log(1 + \c^2)-\frac{9}{4})  +\ldots\Big],
\nn
\end{align}
with the function $f(\chi)$ being completely determined in this instance by the Ward identities.

Under the Lifshitz rescaling \eqref{resc}
\be
\label{tttt}
\left\langle J_{tt}(t,x) J_{tt}(0)\right\rangle \rightarrow \left\langle J_{tt}(\l^z t, \l x) J_{tt}(0)\right\rangle = \frac{1}{\l^2 \l^{2z}}\left\langle J_{tt}(t,x) J_{tt}(0)\right\rangle 
\ee
up to order $\e^4$.

The correlator $\left\langle J_{xx}(t,x) J_{xx}(0)\right\rangle$ agrees with $\left\langle J_{tt}(t,x) J_{tt}(0)\right\rangle$ up to terms without logarithms at order $\e^2$ and therefore it scales as
\begin{align}
\label{xxxx}
\left\langle J_{xx}(t,x) J_{xx}(0)\right\rangle \rightarrow \left\langle J_{xx}(\l^z t, \l x) J_{xx}(0)\right\rangle = \frac{1}{\l^2 \l^{2z}}\left\langle J_{xx}(t,x) J_{xx}(0)\right\rangle.  
\end{align}
Similarly, we can compute
\begin{align}
\label{txttcorr}
 &\left\langle J_{tx}(t,x) J_{tt}(0)\right\rangle = 8 \e^2 \frac{c/2}{(2 \p)^2} \frac{t x (t^4 - 5 t^2 x^2 + 2 x^4)}{(t^2 + x^2)^5} \log(t^2 + x^2) \\& +8(1-\frac{\e^2}{2}) \frac{c/2}{(2 \p)^2} \frac{t x (t^2 -  x^2 )}{(t^2 + x^2)^4} -9\e^2 \frac{c/2}{(2 \p)^2} \frac{tx(t^2 -3x^2)(3t^2-x^2)}{(t^2+x^2)^5} \nn\\
 &+\frac{1}{3} \e^2 \frac{c/2}{(2 \p)^2} \frac{tx}{(t^2 +x^2)^3}+ \ldots. \nn \\
 &= 4\frac{c}{(2 \p)^2}  x^{-(1+z)-2 z} \Big[ \frac{\c(\c^2 -1)}{(\c^2 +1)^4} \nn\\&+\frac{\e^2}{6} \frac{\c(-23 \c^4 + 68 \c^2 -17)}{(\c^2 +1)^5}
+\e^2 \frac{\c(\c^4-5\c^2+2) }{(\c^2+1)^5} \log(1 + \c^2)  +\ldots\Big].  \nn
\end{align}
Under the rescaling \eqref{resc} this transforms as
\be
\label{txtt}
\left\langle J_{tx}(t,x) J_{tt}(0)\right\rangle \rightarrow \left\langle J_{tx}(\l^z t, \l x) J_{tt}(0)\right\rangle = \frac{1}{\l \l^{3z}}\left\langle J_{tx}(t,x) J_{tt}(0)\right\rangle. 
\ee
Recall that $J_{\m\n}$ is not symmetric and therefore $J_{tx}$ and $J_{xt}$ are not equivalent. For example,
\begin{align}
\label{xtttcorr}
 &\left\langle J_{xt}(t,x) J_{tt}(0)\right\rangle 
 = 4\frac{c}{(2 \p)^2}  x^{-2-(1+z)} \Big[ \frac{\c(\c^2 -1)}{(\c^2 +1)^4} \\&+\frac{\e^2}{6} \frac{\c(-23 \c^4 + 68 \c^2 -17)}{(\c^2 +1)^5}
+\e^2 \frac{\c(2\c^4-5\c^2+1) }{(\c^2+1)^5} \log(1 + \c^2)  +\ldots\Big],
\nn
\end{align}
and this scales under \eqref{resc} as
\be
\label{xttt}
\left\langle J_{xt}(t,x) J_{tt}(0)\right\rangle \rightarrow \left\langle J_{xt}(\l^z t, \l x) J_{tt}(0)\right\rangle = \frac{1}{\l^3 \l^{z}}\left\langle J_{xt}(t,x) J_{tt}(0)\right\rangle. 
\ee
From these results  we can now read off the Lifshitz scaling dimensions for the different components of $J$. From \eqref{tttt} and \eqref{xxxx} we see that the scaling dimension of $J_{tt}$ equals that of $J_{xx}$
\be
\D_L (J_{tt}) = \D_L (J_{xx}) = 1+z,
\ee
where the subscript $L$ is used to signify that this is the Lifshitz scaling dimension. Furthermore \eqref{txtt} and \eqref{xttt} imply that
\be
\D_L (J_{tx}) = 2z, \qquad \D_L (J_{xt}) =2.
\ee
These scaling dimensions agree with those derived on general grounds in \cite{Ross:2011gu} and thus our correlators indeed display the expected scaling properties. Note that one interprets $J_{tt}$ and $J_{xt}$ as the energy and momentum densities, respectively, while $J_{tx}$ and $J_{xx}$ are the energy fluxes and momentum fluxes, respectively. Note that in a non-relativistic theory momentum density and energy flux are independent quantities. 

One can derive the remaining correlators using the diffeomorphism Ward identity
\begin{align}
\pa_t \left\langle J_{tt}(t,x)\right\rangle + \pa_x \left\langle J_{tx}(t,x) \right\rangle &=0, \\
\pa_t \left\langle J_{xt}(t,x)\right\rangle + \pa_x \left\langle J_{xx}(t,x) \right\rangle &=0,
\end{align}
just as in the last subsection. For example, using the explicit expressions given above it is straightforward to check that
\be
\pa_x \left\langle J_{tx}(t,x) J_{tt}(0)\right\rangle =- \pa_t \left\langle J_{tt}(t,x)  J_{tt}(0)\right\rangle.
\ee
To display the full structure contained in the correlators it is useful to introduce the notation
\be
\label{fcorr}
\left\langle J_{\m\n}(t,x) J_{\r \s}(0)\right\rangle = x^{- \D_L(J_{\m\n}) - \D_L(J_{\r\s})} f_{\m\n,\r\s}(\c).
\ee
Note that $f_{\m\n,\r\s}(\c)$ is not a tensor but rather a bookkeeping device. One can then translate the diffeomorphism Ward identities into simple differential relations between the different components of $f_{\m\n,\r\s}$. For example
\be
f'_{tt,tt}(\c) = (1+3z)f_{tx,tt}(\c) + z \c f'_{tx,tt}(\c),
\ee
where a prime denotes a derivative with respect to $\c$. It is easy to verify this relation holds using \eqref{ttttcorr} and \eqref{txttcorr}. If the Ward identities do not get modified at higher orders in $\e$ than this and analogous relations for other components would hold to all orders in $\e$!
Explicit expressions for all remaining correlation functions may be found in Appendix \ref{apcorr}.

\section{Dual QFT} \label{sec:QFT}

In previous sections we found that the QFT dual to the Lifshitz geometries with $z=1+\e^2$ is a
specific deformation of a CFT. One might wonder whether such Lifshitz critical points can only
arise in strongly interacting QFTs with a holographic dual or whether there is a general such construction of
Lifshitz invariant theories, irrespectively of whether the theory is strongly or weakly coupled, has a
holographic dual or not. We show in this section that this construction indeed holds in general.
We consider a CFT deformed by a weight $d$ vector ${\cal J}^i$,
\begin{align}
 S = S_{\rm CFT} + \sqrt{2} \e \int d^d x {\cal J}^t.
\end{align}
(The factor of $\sqrt{2}$ is included only for the purpose of comparison with the earlier holographic discussion.)

\subsection{The classical theory}

Let us first discuss the classical theory. After the deformation, the theory still has a conserved energy-momentum tensor
${\cal T}_{ij}$, since the theory is invariant under translations. However, this tensor is not symmetric any longer because the
deformed theory is not Lorentz invariant \cite{Ross:2009ar,Guica:2010sw}.
One can work out ${\cal T}_{ij}$ either as the Noether current corresponding to translations or by coupling the theory to a
vielbein $e_i^{\hat{k}}$  and varying with respect to it (hatted indices correspond to flat tangent directions). In our case the coupling to vielbein is given by
\be
S[e] =  S_{\rm CFT}[e] +  \epsilon \sqrt{2} \int d^dx e e_i{}^{\hat{t}}  J^i,
\ee
and the stress energy tensor is defined by
\be
{\cal T}_i{}^{\hat{k}} = -\frac{1}{e} \frac{\delta S[e]}{\delta e^i_{\hat{k}}}.
\ee
This is a conserved tensor, $\nabla^i {\cal T}_i{}^{\hat{k}}=0$.
In our case, it is given by
\be \label{ST}
{\cal T}_{ij} \equiv {\cal T}_i{}^{\hat{k}} e_{j\hat{k}}= T^{CFT}_{ij} + \sqrt{2} \epsilon
\left(g_{ij} J^{\hat{t}}
+e_i{}^{\hat{t}} J_j - e_{j\hat{k}}  \left(\frac{\delta J^k}{\delta e_{\hat{k}}^i}\right) e_k^{\hat{t}} \right).
\ee
From the energy-momentum tensor at hand we can construct the conserved current corresponding to Lifshitz rescaling. Consider the following current
\be
l_i = {\cal T}_{ij} \xi^{j}
\ee
where $\xi^i$ is the Lifshitz transformation,
\be
\delta x^i = \xi^i, \qquad \xi^0 = z x^0, \quad \xi^a = x^a,  \quad (a=1, \ldots, d-1).
\ee
Taking the divergence of this current we find
\begin{align}
\partial^{i} l_{i} = (\partial^{i} {\cal T}_{ij}) \xi^{j} + {\cal T}_{ij} \partial^{i}\xi^{j},
\end{align}
where we now consider the theory in a flat background.
The first term vanishes due to translational invariance. In the relativistic case,  ${\cal T}_{ij}$ is symmetric, and at this point one
symmetrizes, $\nabla^{i}\xi^{j} \to 1/2 (\partial^{i}\xi^{j}+\partial^{j}\xi^{i})$. Then the second term
vanishes if $\xi^i$ is a Killing vector
(Poincar\'{e} transformations) or it can be made to be proportional to the trace of stress energy tensor, ${\cal T}^i_i$, if $\xi^i$ is a
conformal Killing vector. In the latter case conformal invariance is thus  linked to the tracelessness of ${\cal T}_{ij}$.

In our case,  ${\cal T}_{ij}$ is not symmetric.  However,
\be
\partial_{0}\xi^{0}=z, \qquad \partial_{a}\xi^{b}= \delta_{a}^{b}  \qquad \partial_{a}\xi^{0}= \partial_{0}\xi^{a}=0.
\ee
It follows that the conservation of $l_i$ is equivalent to
\begin{align} \label{Lifinvariance}
0=\nabla^{i} l_{i} = z {\cal T}^t_{t} + {\cal T}^a_a.
\end{align}
We conclude that a non-relativistic theory with a (non-symmetric) stress energy tensor ${\cal T}_{ij}$ is Lifshitz invariant
if ${\cal T}_{ij}$ satisfies the trace condition (\ref{Lifinvariance}).

Taking the trace of (\ref{ST}) we find
\be
{\cal T}^i_i = \sqrt{2} \e \left((d+1) J^{\hat{t}} + (\delta_D J^i) e_i{}^{\hat{t}}\right) =0,
\ee
where in the first equality we used the fact that the stress energy tensor of original CFT is traceless
and in the second, $\delta_D J^i=-(d+1) J^i$, which expresses the fact that $J^i$ is a weight $d$ vector.
Thus, at the classical level we have a non-relativistic $z=1$ Lifshitz theory.

\subsection{Conformal perturbation theory}

We now turn to the quantum theory. We will eventually specialize to the case of a deformation with only the time component participating but we start by considering a more general deformation:
\begin{align}
 S = S_{\rm CFT} + \e \int d^d x A_{(0) i} {J}^i.
\end{align}
Since the deformation is small we can study the theory using conformal perturbation theory.
Let us consider the partition function $Z[\e]$ and expand in $\e$,
\bea
Z[\e]&=&Z_{CFT} - \e \int d^d x A_{(0)i} \langle {J}^i (x)\rangle_{CFT} \label{z} \\
&& \quad + \frac{1}{2} \e^2 \int_{|x-y|>\L} d^dx d^d y A_{(0)i}(x) A_{(0)j}(y)  \langle { J}^i(x) {J}^j(y) \rangle_{CFT} \nn
\eea
where  $\langle\  \rangle_{CFT}$ denotes the computation in the conformal vacuum of the undeformed theory  and $1/\L$ is a UV cut-off. Since CFT 1-point functions vanish,
the  leading non-trivial effect is at order $\e^2$. To compute this we will use the OPE of the vector operators.

The general form of the OPE is
\be
{J}_{i}(x) {J}_j(0) \sim \sum C_{ij}^k \frac{{\cal O}_k}{x^{2d -\Delta_k}},
\ee
where $\Delta_k$ is the dimension of the operator ${\cal O}_k$. Inserting this in (\ref{z}) one finds divergences whenever
\be
\Delta_k \le d.
\ee
To remove them  we will need to renormalize the sources of ${\cal O}_k$.
If we do not have couplings to these operators we have to add them at
this point. The OPE contains the following universal terms
\begin{align} \label{cr-norm}
{J}_{i} (x) {J}_{j} (0) \sim  C_J \frac{I_{ij}}{x^{2d}} +\cdots+ {\cal A}_{ij}{}^{kl} \frac{{\cal T}_{kl}}{x^d} + \ldots,
\end{align}
where
\be
I_{ij}=\delta_{ij} - 2 \frac{x_{i} x_{j}}{x^2}.
\ee
and the overall normalization is correlated
with the normalization of ${J}_{i}$. The OPE coefficient
${\cal A}_{ij}{}^{kl}$ is completely fixed by conformal invariance in $d=2$
while there is a 2-parameter family of coefficients when $d>2$. We will
discuss the two cases in turn. The terms not exhibited are theory specific rather than universal.

\subsubsection{From a relativistic to a  Lifshitz critical point}

Before we move on to discuss in detail the two cases let us explain how the relativistic critical point becomes a Lifshitz
invariant critical point. After the relativistic CFT is deformed there are beta functions and one finds that the dilation Ward identity becomes
\be \label{tii}
\<\mathcal{T}^i_i\>= -\sum_i \beta_i O^i.
\ee
Zeroes of the beta functions will lead to a new relativistic CFT in the IR, since then $\<\mathcal{T}^i_i\>=0$.
As we reviewed above, Lifshitz invariance is characterized by
\be
z\<\mathcal{T}^t_t\> + \<\mathcal{T}^a_a\> =0.
\ee
Thus in order to obtain a Lifshitz invariant fixed point starting from a relativistic one, one of the operators
appearing on the rhs of (\ref{tii}) must be the stress energy tensor and it should have a non-zero beta function such that
\be
\<\mathcal{T}^i_i\> + \frac{1}{2} \beta_g^{ij} \<\mathcal{T}_{ij}\> = z\<\mathcal{T}^t_t\> + \<\mathcal{T}^a_a\>,
\ee
for some $z$.
In other words, this beta function, instead of generating a flow, changes the condition of scale invariance from the relativistic one to a Lifshitz invariant one. If there are beta functions beyond the one for metric in (\ref{tii}) then one needs these to be zero to remain at a fixed point.
We will show in the next subsections that the deformation we consider is indeed of this type.

\subsubsection{The d=2 OPE}

In this section we work in two dimensions, in Euclidean signature, introducing complex coordinates
$z=x_1 + i x_2, \bar{z}=x_1 - i x_2$ \footnote{We use the conventions in \cite{DiFrancesco:1997nk}. In partricular,
$\partial = \frac{1}{2}(\partial_{x_1} - i \partial_{x_2}), \quad \bar{\partial}=\frac{1}{2}(\partial_{x_1} + i \partial_{x_2}), \quad v_z= \frac{1}{2}(v^{x_1} - i v^{x_2}), \quad v_{\bar{z}} = \frac{1}{2}(v^{x_1} + i v^{x_2}), 
g_{z \bar{z}} = \frac{1}{2}, \quad
d^2 z = 2 dx_1 dx_2, \quad \delta^2(z) = \frac{1}{2} \delta(x_1) \delta(x_2)$ 
Note also the useful identity: $\partial \bar{\partial} \log |z|^2 = 2 \p \delta^2 (z)$.}. The vector operator $J\equiv J_z$
has dimension $(h_J,\bar{h}_J)$, while $\bar{J} \equiv J_{\bar{z}}$ has dimension
$({h}_{\bar{J}},\bar{h}_{\bar{J}})$.
The stress energy tensor is defined as usual by $\mathcal{T}_{\m \n} = \frac{2}{\sqrt{g}}\frac{\delta S}{\delta g^{\m\n}}$ and (following \cite{DiFrancesco:1997nk}) we also define
\begin{align} \label{defT}
T(z) = -2 \p  \mathcal{T}_{zz}
\end{align}
with a similar formula for the anti-holomorphic part.
$T$ has the standard normalization for a $2d$ CFT while the normalization of the holographic stress energy tensor is that of $\mathcal{T}$.

Recall that
\begin{align}
\label{TJOPE}
 T(z) J(w) &\sim \frac{h_J J(w)}{(z-w)^2} + \frac{\partial J(w)}{(z-w)} + \ldots, \\
  T(z) T(w) &\sim \frac{c}{2(z-w)^4} + \ldots
\end{align}
with obvious generalizations for antiholomorphic components of the vector and stress-energy tensor. The central charge of the CFT is $c$.

On general grounds the OPE $J(z) J(w)$ takes the following form:
\begin{align}
 J(z) J(w) &\sim -\frac{C_J/2}{(z-w)^3 (\bar{z}-\bar{w})} + k \frac{T}{(z-w) (\bar{z}-\bar{w})} + \ldots \label{JJOPE}  \\
 \bar{J}(z) \bar{J}(w) &\sim -\frac{C_J/2}{(z-w) (\bar{z}-\bar{w})^3} + k \frac{\bar{T}}{(z-w) (\bar{z}-\bar{w})} + \ldots \\
 {J}(z) \bar{J}(w) &\sim -c_1 {C_J} \frac{\delta^2(z-w, \bar{z} - \bar{w})}{(z-w) (\bar{z}-\bar{w})}+\ldots
\end{align}
We now consider how to determine the constant $k$. We consider the $3$-point function $\left\langle T(z) J(z_1) J(z_2) \right\rangle $ in the limit when (first) $z_2$ goes to $0$ and (then) $z_1$ approaches $z_2$. This correlation function can be determined in two different ways. First we can exploit the $J(z_1) J(z_2)$ OPE directly in the correlator (neglecting a possible trace anomaly, which does not play a role)
\begin{align}
 \left\langle T(z) J(z_1) J(z_2) \right\rangle = \left\langle T(z) k \frac{T(z_2)}{z_1 \bar{z}_1} \right\rangle = k \frac{c}{2} \frac{1}{z_1 \bar{z}_1} \frac{1}{z^4},
\label{1comp}
\end{align}
where we set $z_2=0$. Note that in general the $J(z_1) J(z_2)$ OPE contains also descendants of $T$. But such terms have less singular behaviour when $z_1$ goes to $0$ and it suffices to keep the most singular term only.

Another way to compute this limit of the correlator is to consider an inversion with respect to $z_2$ (or a point very close to it)
\begin{align}
 x'^i = \frac{x^i}{x^2}.
\label{inversion}
\end{align}
This way we may send $z_2$ to infinity and then apply the short-distance expansion \eqref{TJOPE} to calculate $T(z') J(z'_1)$. The inversion \eqref{inversion} corresponds to a local dilatation and a local rotation, 
such that
\begin{align}
 \left\langle T(z) J(z_1) J(z_2) \right\rangle = \frac{1}{(z \bar{z})^2} \frac{1}{(z_1 \bar{z}_1)^2} \frac{1}{(z_2 \bar{z}_2)^2} \frac{\bar{z}^2}{z^2} \frac{\bar{z}_1}{z_1} \frac{\bar{z}_2}{z_2}  \left\langle \bar{T}(z') \bar{J}(z'_1) \bar{J}(z'_2) \right\rangle.
\end{align}
The correlation function of the right-hand side then becomes
\begin{align}
\left\langle \bar{T}(z') \bar{J}(z'_1) \bar{J}(z'_2) \right\rangle &= \left\langle (\frac{\bar{h}_{\bar{J}} \bar{J}(z'_1)}{(\bar{z}'-\bar{z}'_1)^2} + \frac{\bar{\partial} \bar{J}(z'_1)}{(\bar{z}'-\bar{z}'_1)}) \bar{J}(z'_2) \right\rangle \\
&= -\frac{C_J \bar{h}_{\bar{J}}}{2} \frac{1}{(\bar{z}'-\bar{z}'_1)^2 (\bar{z}'_1-\bar{z}'_2)^3 ({z}'_1-{z}'_2)}+\ldots. \nn
\end{align}
In the limit when $z_1$ goes to $0$ we obtain
\begin{align}
 \left\langle T(z) J(z_1) J(z_2) \right\rangle = -\frac{C_J \bar{h}_{\bar{J}}}{2}  \frac{1}{z_1 \bar{z}_1} \frac{1}{z^4}.
\label{2comp}
\end{align}
Comparing \eqref{1comp} to \eqref{2comp} we obtain the relation
\begin{align}
k c = -{C_J \bar{h}_{\bar{J}}} = -{C_J {h}_{{J}}}.
\label{kcrelation}
\end{align}
A particular example of such a deformation is given by combining free boson and free fermion CFTs (such that $c = \bar{c} = 3/2$). In such a theory the vector operator $J_{\mu} = i \partial_{\m} X \psi \bar{\psi}$ of the correct dimension and using the standard free field OPEs
\begin{align}
& X(z) X(w) \sim - \frac{1}{4 \p} \ln |z-w|^2 + \ldots, \\
& \psi (z) \psi(w) \sim \frac{1}{2 \p}\frac{1}{z-w} + \ldots.
\end{align}
it is straightforward to check that $C_J=\frac{1}{(2 \p)^3}$ and $k=-\frac{1}{(2 \p)^3}$ in agreement with \eqref{kcrelation}.

\subsubsection{Conformal perturbation theory in d=2}
Next we use the general OPEs \eqref{JJOPE} in order to obtain the beta function and anomaly in the deformed theory. This will allow us to reproduce the structure of our three-dimensional gravity results.

Consider a deformation of the CFT by a term of the form $\e \int d^2 x A_{\m} J^{\m} = \e \int d^2 z(\bar{A} J + A \bar{J})$, i.e.
\begin{align}
S = S_{\text{CFT}} + \e\int d^2 z(\bar{A} J + A \bar{J}).
\end{align}
Then $J$ has dimension $(3/2,1/2)$ and $\bar{J}$ has dimension $(1/2,3/2)$ in the case of interest.

Expanding $\exp[-\e \int d^2 z(\bar{A} J + A \bar{J})]$ to the second order in $A$ and using the OPEs \eqref{JJOPE} we get
\begin{align}
\label{genfunct}
 &\exp[-\e\int d^2 z(\bar{A} J + A \bar{J})] \sim 1 - \e\int d^2 z(\bar{A} J + A \bar{J}) \\&-\frac{\e^2}{2} C_J \int d^2 z_1 d^2 z_2 \Bigg(c_1 \Big({A} (z_1) \bar{A}(z_2) + \bar{A} (z_1) {A}(z_2)\Big) \frac{\delta^2(z_{12} \bar{z}_{12})}{z_{12} \bar{z}_{12}} \nn\\&+ \bar{A} (z_1) \bar{A}(z_2) \Big[\frac{1}{2 z_{12}^3 \bar{z}_{12}} + \frac{3}{2 c} \frac{T(z_2)}{z_{12} \bar{z}_{12}} \Big] +{A} (z_1) {A}(z_2)  \Big[\frac{1}{2 z_{12} \bar{z}_{12}^3} +\frac{3}{2 \bar{c}} \frac{\bar{T}(z_2)}{z_{12} \bar{z}_{12}} \Big] \Bigg). \nn
\end{align}
Let us take a closer look at possible divergences. All the $z_2$ integrals can be explicitly evaluated in polar coordinates, e.g.
\begin{align}
 \int d^2 z_1 d^2 z_2 \frac{{A} (z_1) {A}(z_2)}{z_{12} \bar{z}_{12}^3} = 2 \p \log (\L^{-1})  \int d^2 z_1 {A} (z_1) \bar{\partial}^2 A(z_1),
\label{log1}
\end{align}
where $\L^{-1}<<1$ is a UV cutoff. The cutoff introduces a scale and thus breaks Weyl invariance, and the logarithmic divergence is removed by a logarithmic counterterm. Noting that $A \bar{\partial}^2 A + \bar{A} {\partial}^2 \bar{A}= \frac{1}{16} F_{ij} F^{ij} - \frac{1}{8} (\partial_i A^i)^2 +\text{total derivative}$ we see that it precisely mimics \eqref{logct}.

Another divergent term arises from
\begin{align}
 &\int d^2 z_1 d^2 z_2 \frac{{A} (z_1) {A}(z_2)}{z_{12} \bar{z}_{12}} \bar{T}(z_2) = 4\p \log (\L^{-1}) \int d^2 z_1 {A} (z_1) {A}(z_1)\bar{T}(z_1).
\label{log2}
\end{align}
This involves the stress-energy tensor and thus renormalizes the metric ($\delta S = \frac{\sqrt{g}}{2} T_{\m\n} \delta g^{\m\n}$; recall that $T=-2\p T_{zz}$)
\begin{align}
 g^{\bar{z} \bar{z}} \rightarrow g^{\bar{z} \bar{z}}_{\text{R}}=g^{\bar{z} \bar{z}}-{16 \p^2} \frac{3 C_J}{2 c} \log (\L^{-1}) A_z A_z.
\end{align}
Equivalently
\begin{align}
\frac{\partial g_{\text{R}zz}}{\partial \log (\L^{-1})}=  {4 \p^2} \frac{3 C_J}{2 c} A_z A_z {= A_z A_z} 
\end{align}
and similarly
\begin{align}
\frac{\partial g_{\text{R}\bar{z}\bar{z}}}{\partial \log (\L^{-1})}= {4 \p^2} \frac{3 C_J}{2 c} A_{\bar{z}} A_{\bar{z}} {= A_{\bar{z}} A_{\bar{z}}} ,
\end{align}
{where we used the relation $C_J = c/(6 \p^2)$ which follows from \eqref{J2ptfn}.} In the gravity computation this renormalization arose from the $h_{[2](0)}$ correction in $2$ dimensions. Recall that $h_{[2](0)ij} = - A_{i} A_{j} + \frac{1}{2}A_{k}A^{k} g_{[0](0)ij}$. Using 
\begin{align}
h_{[2](0)zz} &{=}  A_z A_z = \frac{1}{4}(A^t A^t - A^x A^x + 2 i A^x A^t); \\
h_{[2](0)\bar{z} \bar{z}} &{=} A_{\bar{z}} A_{\bar{z}} =\frac{1}{4}(A^t A^t - A^x A^x - 2 i A^x A^t); \\
h_{[2](0)tt} &= h_{[2](0)zz} + h_{[2](0)\bar{z} \bar{z}} = - h_{[2](0)xx} {=}  \frac{1}{2}(A_t A_t - A_x A_x).
\end{align}
If we want to compare
with the gravitational results we should analytically continue to Lorentzian signature $A_{t} A_{t} \rightarrow - A_{\t} A_{\t}$ and $h_{[2](0)tt} \rightarrow -h_{[2](0) \t \t}$. Then the CFT expression for $h_{[2](0)}$ agrees exactly with the gravity computation.
Switching on only the deformation by the time-component of the vector leads to $z=1+\epsilon^2$, in agreement with our bulk computation.

The most leading divergence comes from
\begin{align}
 \int d^2 z_1 d^2 z_2 {A}(z_1) \bar{A} (z_2) \frac{\delta^2(z_{12},\bar{z}_{12})}{z_{12} \bar{z}_{12}} = 2\int_{\L^{-1}}d\r \frac{\delta(\r)}{\r^2} \int d^2 z_1 A(z_1) \bar{A}(z_1).
\end{align}
This divergence is removed by a local counterterm which is the counterpart of \eqref{lct}.

The generating functional of connected diagrams transforms under Weyl variations as  \cite{Osborn:1991gm}
\begin{align}
\delta_W W =\frac{d W}{d \log(\L)}= \frac{\partial W}{\partial \log(\L)}+\sum_i \b_i \mathcal{O}^i + {a},
\end{align}
where $\b_i$ are the beta functions for the operators coupled to $\mathcal{O}^i$ and ${a}$ is the trace anomaly (not to be confused with $\mathcal{A}$). In our case there are non-vanishing beta functions for the metric $g$. 

On the other hand \cite{Osborn:1991gm}
\begin{align}
\frac{\partial W}{\partial \log(\L)}=-\langle T_i^i \rangle + \langle J^i \rangle A_i.
\end{align}
Comparing with \eqref{Ward} this gives us an interpretation of the non-local term $A^i t_{ij} A^j$ appearing in $\mathcal{A}$: it comes from beta functions!

We can compute $\frac{d W}{d \log(\L)}$ directly from the renormalized action. $\mathcal{A}$ arises essentially from the logarithmic divergences \eqref{log1} and \eqref{log2}, which combine to give a total logarithmic divergence proportional to 
\begin{align}
\Big( A \bar{\partial}^2 A + \bar{A} {\partial}^2 \bar{A} \Big) + 4 \frac{3}{2 c}(\bar{A} T \bar{A} + A \bar{T} A).
\end{align}
Note that
\begin{align}
 A \bar{\partial}^2 A + \bar{A} {\partial}^2 \bar{A}= \frac{1}{16} F_{ij} F^{ij} - \frac{1}{8} (\partial_i A^i)^2 +\text{total derivative}
\end{align}
and
\begin{align}
\bar{A} T \bar{A} + A \bar{T} A = \frac{1}{4} A^i T_{ij} A^j.
\end{align}
Thus the logarithmic divergence (which is equal in this case to the anomaly) is
\begin{align}
 \Big(\frac{1}{16} F_{ij} F^{ij} - \frac{1}{8} (\partial_i A^i)^2 \Big) +\frac{3}{2 c} A^i T_{ij} A^j.
\end{align}
To compare to the gravitational computation recall that Newton's constant is related to the central charge of the underlying CFT through $c=\frac{3}{2 G_3}$. We rewrite our holographic anomaly \eqref{tranomaly} as
\begin{align}
\mathcal{A} = \frac{1}{2}A^i_{(0)} \langle T_{ij} \rangle_{[0]} A^j_{(0)} - \frac{c}{24 \p} \Big(\frac{1}{4} F_{(0)ij}F_{(0)}^{ij}-\frac{1}{2}(\nabla_i A^i_{(0)})^2 \Big),
\end{align}
where we omitted the curvature term as we cannot see it in our CFT computation because we are working in a flat background.
Recalling that $T_{\text{CFT}} = - 2 \p T_{\text{bulk}}$ we see that the gravity and CFT computations  indeed produce exactly the same Weyl anomaly. {Recalling that in our holographic model $C_J = 1/(4 \p^2 G_3)$ one can check that even the overall coefficient of the anomaly agrees with the gravity computation.}

\subsubsection{d>2} \label{TJJsection}

In this section we adapt the discussion of \cite{Osborn:1993cr} to the case of the $3$-point function $\left\langle T_{\m\n}(x_1) J_{\s}(x_2) J_{\r}(x_3) \right\rangle$, where $J$ is a vector field of dimension $\D$ (equivalent results can be obtained using the embedding formalism \cite{Costa:2011mg}). Here we will assume that there is a unique spin-$2$ conserved current i.e. the stress-energy tensor is unique. Our goal is to derive the general OPE and compute the beta function for the metric. 

Our starting point is the following expression \cite{Osborn:1993cr} for the $3$-point function under consideration:
\begin{align}
\label{TJJcorr}
\left\langle T_{\m\n}(x_1) J_{\s}(x_2) J_{\r}(x_3) \right\rangle &= \frac{1}{x_{12}^d x_{13}^d x_{23}^{2\D - d}} \mathcal{I}_{\m\n,\g\delta}(x_{13}) I_{\s\a}(x_{23}) \tilde{t}_{\g\delta\a\r}(X_{12})\\
&=\frac{1}{x_{12}^d x_{13}^d x_{23}^{2\D - d}}  I_{\s\a}(x_{13}) {I}_{\r\b}(x_{23}) {t}_{\m\n\a\b}(X_{23}),\nn
\end{align}
where $x_{ij} = x_i - x_j$, $I_{\m\n}(x) = \delta_{\m\n} - 2 \frac{x_{\m} x_{\n}}{x^2} $, $X_{12} = \frac{x_{13}}{x_{13}^2} - \frac{x_{23}}{x_{23}^2}$ and
\begin{align}
\mathcal{I}_{\m\n,\s\r}(x) = \frac{1}{2}(I_{\m\s}(x) I_{\n\r}(x) + I_{\n\s}(x) I_{\m\r}(x)) - \frac{1}{d}\delta_{\m\n} \delta_{\s\r}.
\end{align}
The tensors $t$ and $\tilde{t}$ are homogeneous of degree zero in $X$ and they satisfy
\begin{align}
\tilde{t}_{\m\n\s\r}(X) = I_{\s\a}(X) t_{\m\n\a\r}(X), \quad t_{\m\n\s\r} = t_{\n\m\s\r}=t_{\m\n\r\s}, \quad t_{\m\m\s\r}=0.
\end{align}
We can write the OPE of $T_{\m\n}$ with $J_{\r}$ in the form
\begin{align}
\label{TJope}
T_{\m\n}(x_1)J_{\r}(x_2) \sim A_{\m\n\r\s}(x_{12}) J^{\s}(x_2) + B_{\m\n\r\l\s}(x_{12}) \partial^{\l}J^{\s}(x_2)+\ldots.
\end{align}
Using the methods of \cite{Osborn:1993cr} one can show that
\be
\tilde{t}_{\g\delta\a\r}(X_{12})=C_J \frac{x^d_{12}}{x^d_{13} x^d_{23}} A_{\g\delta\a\r}(X_{12}).
\ee
The OPE coefficient $A_{\m\n\s\r}$ must be traceless and symmetric in the first two indices and it must satisfy $\partial^{\m}A_{\m\n\s\r}=0$. Furthermore, $I_{\s\l}A_{\m\n\l\r}$ must be symmetric in $\s$ and $\r$ (this can be shown by multiplying the OPE \eqref{TJope} with $J_{\l}$ and taking expectation value on both sides). This fixes its form to be
\begin{align}
A_{\m\n\s\r}(x) =&\Big[ (a + db) h^1_{\m\n}(x) g_{\s\r} + (d^2-4) b h^1_{\m\n}(x) h^1_{\s\r}(x) \\&+ b (h^2_{\m\n\s\r}(x) - h^3_{\m\n\s\r}(x)) + e \tilde{h}_{\m\n\s\r}(x) \Big]\frac{1}{x^d}\nn
\label{AOP}
\end{align}
with so far undetermined constants $a$, $b$, $e$. We have introduced the following notation from \cite{Osborn:1993cr}
\begin{align}
h^1_{\m\n}(x) &= \frac{x_{\m} x_{\n}}{x^2} - \frac{1}{d} g_{\m\n}, \\
h^2_{\m\n\s\r}(x) &= \frac{x_{\m} x_{\s}}{x^2} g_{\n\r} + (\m \leftrightarrow \n, \s \leftrightarrow \r) - \frac{4}{d} \frac{x_{\m} x_{\n}}{x^2} g_{\s\r} - \frac{4}{d} \frac{x_{\s} x_{\r}}{x^2} g_{\m\n} + \frac{4}{d^2} g_{\m\n} g_{\s\r}, \nn\\
h^{3}_{\m\n\s\r} (x)& = g_{\m\s} g_{\n\r} + g_{\m\r} g_{\n\s} - \frac{2}{d} g_{\m\n} g_{\s\r},\nn\\
\tilde{h}_{\m\n\s\r}(x) &= \frac{x_{\m} x_{\s}}{x^2} g_{\n\r} + \frac{x_{\n} x_{\s}}{x^2} g_{\m\r} - \frac{x_{\m} x_{\r}}{x^2} g_{\n\s} - \frac{x_{\n} x_{\r}}{x^2} g_{\m\s}. \nn
\end{align}

Under conformal transformations the transformation of the current is given by the integral over the sphere \cite{Cardy:1987dg}
\begin{align}
\delta J_{\s}(0) = -\int_{x=\e} \a^{\m}(x) T_{\m\n}(x) x^{\n} x^{d-2}J_{\s}(0)d\O
\label{transf}
\end{align}
where $d\O$ is normalized such that $\int d\O =1$.
If we now consider a dilatation $\a^{\m} = \a x^{\m}$ under which $\delta J_{\s}(0) = \D \a J_{\s}(0)$ we get
\begin{align}
\D J_{\s}(0) &= -\int x^{\m} x^{\n} x^{d-2} A_{\m\n\s\r}(x) J^{\r}(0) d\O \\
&= -\int \frac{1}{x^2} \Big((d+1) (d-2) b x_{\r} x_{\s} + \frac{(d-1)a + 2b}{d} x^2 g_{\r\s} \Big)J^{\r}(0) d\O.\nn
\end{align}
(this is equivalent to the Ward identity at coincident points).
Evaluating the integrals we obtain an additional relation 
\begin{equation}
a+db = -\frac{d \D}{d-1}.
 \end{equation}
 The leading coefficient of the OPE is not restricted by the special conformal and translation transformations because their contribution to the integral in \eqref{transf} vanishes (the integrand is odd in $x_i$). Thus, the leading term in the OPE of $T_{\m\n}$ with $J_{\r}$ is fixed up to two independent coefficients.

As an aside, we note that the two dimensional case is special. In his case, the scaling of operators is characterized by two parameters, $h_J$ and $\bar{h}_J$,
instead of one (the overall conformal dimension $\Delta = h_J + \bar{h}_J=d$). In two flat dimensions \eqref{AOP} becomes
\begin{align}
A_{\m\n\s\r}(x) &=\Big[ (a + 2b) h^1_{\m\n}(x) g_{\s\r}  + e \tilde{h}_{\m\n\s\r}(x) \Big]\frac{1}{x^2} \\&= \Big[-4 h^1_{\m\n}(x) g_{\s\r}  + e \tilde{h}_{\m\n\s\r}(x) \Big]\frac{1}{x^2},\nn
\label{Azz}
\end{align}
since $h^2_{\m\n\s\r}(x) - h^3_{\m\n\s\r}(x)$ vanishes identically. The only independent constant $e$ is determined entirely by the holomorphic weight $h_J$.
One can see this by considering the OPE \eqref{Azz} in complex coordinates:
\begin{align}
T_{zz} J_z \sim \frac{1}{z^2}(-1+\frac{e}{2}) J_z,
\end{align}
where we have used $h^1_{zz} = \frac{\bar{z}}{4 z}$, $h^1_{\bar{z} \bar{z}} = \frac{{z}}{4 \bar{z}}$, $\tilde{h}_{z z z \bar{z} }  = \frac{{z}}{4 \bar{z}}$.

In two dimensions the Ward identity \eqref{transf} can be rewritten as
\begin{align}
\delta  J_z (0) =  \frac{1}{2 \p i}\int_\G [v(z) T_{zz}(z) - \bar{v}(\bar{z}) \bar{T}_{\bar{z} \bar{z}}(\bar{z})] J_z(z),
\end{align}
where $\G$ is some contour around the origin. Using a holomorphic rescaling $v(z) = z$, $\bar{v}(\bar{z}) =0$, under which $\delta  J_z (0)  = h_J J_z (0) $, we identify
\begin{align}
e = 2 (h_J + 1).
\end{align}

Going back to \eqref{TJJcorr} we can determine part of the $JJ$ OPE. Recall that
\begin{align}
\left\langle T_{\g\delta}(x_2) T_{\m\n}(x_1)\right\rangle = \frac{C_T}{x_1^{2d}} \mathcal{I}_{\g\delta,\m\n}(x_1),
\end{align}
where $C_T$ is a constant determining the overall scale of the correlator; it is an analog of the central charge. Evaluating $\left\langle T J J\right\rangle $ by first using the $JJ$ OPE and comparing it then to \eqref{TJJcorr} we deduce that
\begin{align}
J_{\m}(x) J_{\n}(0) \sim C_J\frac{I_{\m\n}(x)}{x^{2d}} + \ldots +\frac{C_J}{C_T} I_{\m\a}(x) A_{\g\delta \a\n}(x) T_{\g\delta}(0)+\ldots.
\label{3dJJOPE}
\end{align}
Note however that $A_{\g\delta \a\n}(x)$ is traceless in $\g$, $\delta$.

From this OPE we can immediately derive the leading divergence in the partition function
\begin{align}
&\frac{1}{2}\int d^dx d^dy A^{\m}(y) A^{\n}(x)  J_{\m}(y) J_{\n}(x) \\
&= \frac{C_J}{2} \int d^dx d^dy (A^{\m}(x) + \ldots) A^{\n}(x) \frac{I_{\m\n}(y-x)}{(y-x)^{2d}}+\ldots \nn\\
&= \frac{C_J}{2} \frac{(d-2)\L^d}{d^2} \int d^dx  A_{\m}(x) A^{\m}(x) + \ldots \quad (d>2). \nn
\end{align}
In $d=2$ leading divergence equals $\frac{C_J}{2}\L^2 \int d^2x  A_{\m}(x) A^{\m}(x)$. This divergence can be cancelled by the obvious local counterterm. 

The OPE \eqref{3dJJOPE} also allows us to compute the beta function for the background metric. To this end we expand the deformed action up to second order in the deformation
\begin{align}
&\int d^dx d^dy A^{\m}(y) A^{\n}(x)  J_{\m}(y) J_{\n}(x) \\
&= \ldots + \frac{C_J}{C_T} \int d^dx d^dy (A^{\m}(x) + \ldots) A^{\n}(x) I_{\m\a}(y-x) A_{\g\delta \a\n}(y-x) T_{\g\delta}(x)+\ldots. \nn
\end{align}
The logarithmic divergence comes from the $y$-integral
\begin{align}
&\int  d^dy I_{\m\a}(y) A_{\g\delta \a\n}(y) \\
=&\int d^dy \frac{1}{y^d} \Big[-\frac{d \D}{d-1} h^1_{\g\delta}(y)(\frac{d-2}{d}g_{\m\n} - 2 h^1_{\m\n}(y)) \nn\\&+ b(d^2-4) h^1_{\g\delta} (\frac{2-d}{d}h^1_{\m\n} + 2 \frac{1-d}{d^2}g_{\m\n}) + e(-h^2_{\g\delta\m\n} + 4 h^1_{\g\delta}h^1_{\m\n})\nn\\
&+b\Big( \frac{4 (2-d)}{d} h^1_{\g\delta} h^1_{\m\n} + \frac{4(2-d)}{d^2}h^1_{\g\delta} g_{\m\n} + h^2_{\g\delta \m\n} - h^3_{\g\delta\m\n} \Big)  \Big].\nn
\end{align}
Using
\begin{equation}
\int d\Omega \frac{y_i y_j}{y^2} = \frac{g_{ij}}{d}, \qquad \int d\Omega \frac{y_i y_j y_k y_l}{y^4} = \frac{g_{ij}g_{kl}+g_{ik}g_{jl}+g_{il}g_{kj}}{d(d+2)},
\end{equation}
this integral can be evaluated to give
\begin{align}
&\int  d^dy I_{\m\a}(y) A_{\g\delta \a\n}(y) = \log \L \frac{2}{d+2}\Big( \frac{\D}{d-1} + b\frac{(d-2)(d+1)}{d}-e \Big) \P_{\g\delta\m\n},
\end{align}
where $\P_{\g\delta\m\n} = g_{\g\m}g_{\delta\n} + g_{\g\n}g_{\delta\m} - \frac{2}{d}g_{\m\n}g_{\g\delta}$ is the projector on symmetric traceless part.
Thus the beta function is
\begin{align}
\beta_{\m\n} = 2\frac{2}{d+2} \frac{C_J}{C_T}\Big( \frac{\D}{d-1} + b\frac{(d-2)(d+1)}{d}-e \Big) (A_{\m} A_{\n} - \frac{1}{d} A_{\l} A^{\l}g_{\m\n}).
\label{betafn}
\end{align}
Note that it is traceless because $A_{\g\delta \a\n}$ is traceless in its first two indices by construction (this just reflects the fact that the stress energy tensor it multiplies is traceless). Thus we indeed obtain the expected beta function.

\subsection{Three-dimensional examples}

We now turn to specific examples of three dimensional field theories.

\subsubsection{Example I}

Consider a theory of two free scalar fields $\phi_1$ and $\phi_2$ with the stress energy tensor
\begin{align}
T_{\m\n} = \partial_{\m} \phi_1 \partial_{\n} \phi_1  - \frac{1}{8}\Big( \pa_{\m} \pa_{\n}+ \delta_{\m\n} \pa^2 \Big) \phi^2_1 + (1 \leftrightarrow 2).
\end{align}
The propagators are $\langle \phi_1(x) \phi_1(0) \rangle =\langle\phi_2(x) \phi_2(0)\rangle = \frac{1}{S_3} \frac{1}{x} $, where $S_3$ is the volume of the $3$-sphere. For notational convenience we set $S_3 =1$. The constant $C_T=3/2$ in this theory.

We can construct a dimension three vector operator as
\begin{align}
J_{\m} = (\phi_1^2 - \phi_2^2) (\phi_2 \partial_{\m} \phi_1 -  \phi_1 \partial_{\m} \phi_2).
\end{align}
It is straightforward to check that this vector is a conformal primary operator of dimension $3$, i.e. it transforms according to
\be
\label{trrule}
{J'}^{\m}(x') = J^{\frac{\Delta-1}{d}} \frac{\pa {x'}^{\m}}{\pa x^{\n}}{J}^{\n}(x)
\ee
with $\D = 3$ and where $J$ is the Jacobian of the coordinate transformation.

Now we compute the $A_{\m\n\s\r}$ coefficient in the OPE of $T_{\m\n}$ with $J_{\s}$, see section \ref{TJJsection}). The result is
\begin{align}
&T_{\m\n}(x) J_{\s}(0) \sim\\ &\cdots + \frac{3}{4} \Big[-5 \frac{x_{\m} x_{\n} x_{\s} x_{\r}}{x^7} J_{\r}(0)  - 3 \frac{x_{\n}x_{\s}J_{\m}(0) + x_{\m} x_{\s}J_{\n}(0)  + x_{\m} x_{\n} J_{\s}(0)}{x^5} \nn\\&+ 3 \frac{\delta_{\m\n} x_{\s} x_{\r}}{x^5}J_{\r}(0) + \frac{x_{\r}}{x^5} \Big(\delta_{\n\s} x_{\m} + \delta_{\m\s} x_{\n}  \Big)J_{\r}(0) \nn\\
& + \frac{1}{x^3} \Big( \delta_{\n\s} J_{\m}(0) + \delta_{\m\s} J_{\n}(0)+ \frac{1}{3} \delta_{\m\n} J_{\s}(0) \Big) \Big]+\cdots,\nn
\end{align}
where we omitted contributions which are not proportional to $J_{\m}$. This form matches \eqref{AOP} with coefficients $b = -3/4$ and $e=-3/2$.

As a consistency check we compute
\begin{align}
J_{\m}(x) J_{\n} (0) & \sim \frac{8}{x^6} I_{\m\n}(x) + \frac{8}{x^5} \Big(3 \delta_{\m\n} - 7 \frac{x_{\m} x_{\n}}{x^2} \Big) \phi_1^2(0) \\
&+\frac{8}{x^5} \phi_1(0) \Big(x_{\m} \pa_{\n}\phi_1(0) - x_{\n} \pa_{\m} \phi_1(0) \Big) \nn\\
&+ \frac{8}{x^5} \Big(3 \delta_{\m\n} - 7 \frac{x_{\m} x_{\n}}{x^2} \Big)x_{\s}  \phi_1(0) \pa_{\s} \phi_1(0) \nn\\
&+ \frac{8}{x^3} \Big[\pa_{\m} \phi_1(0) \pa_{\n} \phi_1(0) + \frac{x_{\s}}{x^2} \Big(x_{\m} \pa_{\s}\phi_1(0) \pa_{\n} \phi_1(0) +({\m} \leftrightarrow {\n})\Big) \nn\\
&-\frac{1}{2}\frac{x_{\r} x_{\s}}{x^2}\Big(3 \delta_{\m\n} - 7 \frac{x_{\m} x_{\n}}{x^2} \Big)\pa_{\s} \phi_1(0) \pa_{\r} \phi_1(0)   \Big] \nn\\
& + \frac{8}{x^3} \Big[ -\frac{1}{2} \frac{x_{\s} x_{\n}}{x^2} \pa_{\m} \pa_{\s}\phi_1^2(0) +\frac{1}{4}\frac{x_{\r} x_{\s}}{x^2}\Big(3 \delta_{\m\n} - 7 \frac{x_{\m} x_{\n}}{x^2} \Big)\pa_{\s} \pa_{\r} \phi_1^2(0)   \Big] \nn\\&+ (1\rightarrow 2) + \ldots \nn
\end{align}
First, we can read off $C_J = 16$. Secondly, we note that the last line includes descendants of $\phi_1^2$. Since the $2$-point function $\left\langle T_{\m\n}(x) \phi_1^2(y)\right\rangle$ vanishes, these terms do not contribute to the $3$-point function $\left\langle T_{\m\n}(x_1) J_{\s}(x_2) J_{\r} (x_3)\right\rangle$ and therefore they do not contribute to $A_{\g\delta\a\r}$ coefficient in the $3$-point function or OPE. The remaining terms which are quadratic in derivatives and fields (fourth and fifth lines) are what we are really interested in since these should be equal to $I_{\m\a}(x) A_{\g\delta\a\n}(x) T_{\g\delta}(0)$. We find that
\begin{align}
&I_{\m\a}(x) A_{\g\delta\a\n}(x) T_{\g\delta}(0) \\&= \frac{8}{x^3} \Big[\pa_{\m} \phi_1(0) \pa_{\n} \phi_1(0) + \frac{x_{\s}}{x^2} \Big(x_{\m} \pa_{\s}\phi_1(0) \pa_{\n} \phi_1(0) +({\m} \leftrightarrow {\n})\Big) \nn \\
&-\frac{1}{2}\frac{x_{\r} x_{\s}}{x^2}\Big(3 \delta_{\m\n} - 7 \frac{x_{\m} x_{\n}}{x^2} \Big)\pa_{\s} \phi_1(0) \pa_{\r} \phi_1(0)   \Big] + \text{descendants of} \;\;\; \phi_1^2 \nn\\
&+(1\rightarrow 2). \nn
\end{align}
We conclude that up to unimportant descendant fields this theory reproduces \eqref{AOP} with coefficients $b = -3/4$ and $e=-3/2$. This immediately gives the beta function for the background metric field according to \eqref{betafn}.

\subsubsection{Example II}
In our second example we consider the theory of one free real scalar $\phi$ and one free real fermion $\psi$. The stress-energy tensor is given by
\begin{align}
T_{\m\n} =T_{\m\n}^{\phi} + T_{\m\n}^{\psi},
\end{align}
with
\begin{align}
T_{\m\n}^{\phi}&=\partial_{\m} \phi \partial_{\n} \phi  - \frac{1}{8}\Big( \pa_{\m} \pa_{\n}+ \delta_{\m\n} \pa^2 \Big) \phi^2, \\
T_{\m\n}^{\psi} &=\frac{1}{2} \bar{\psi}(\g_{\m} \overset\leftrightarrow{\pa}_{\n} + \g_{\n} \overset\leftrightarrow{\pa}_{\m})\psi.
\end{align}
Note that any linear combination of $T_{\m\n}^{\phi}$ and $T_{\m\n}^{\psi}$ is still a conserved current. which leads to some complications as we see below.
We can construct a conformal primary $\D=3$ vector
\be
J_{\m} = \phi^2 \bar{\psi} \g_{\m} \psi.
\ee
The basic propagators are
\be
\left\langle \phi(x) \phi(0)\right\rangle = \frac{1}{x}, \quad \left\langle \psi(x) \bar{\psi}(0)\right\rangle = \frac{\g \cdot x}{x^3},
\ee
where the Dirac gamma matrices satisfy as usual $\{\g_{\m}, \g_{\n}\} = 2 g_{\m\n}$. Using this information it is straightforward to obtain
\be
C_J =4, \quad C_T^{\phi} = 3/2, \quad C_T^{\psi} = -3.
\ee
We compute the relevant term in the OPE as
\begin{align}
&T_{\m\n}(x) J_{\s}(0) \sim  \\
& \sim \ldots + \frac{1}{x^3} \Big(\frac{3}{2} g_{\m\n} J_{\s} - \frac{9}{2} \frac{x_{\m} x_{\n}}{x^2} J_{\s} + \frac{3}{2} \frac{x_{\r}}{x^2}(g_{\m\s} x_{\n} + g_{\n\s} x_{\m}) J_{\r} \nn\\&- \frac{3}{2}\frac{x_{\s}}{x^2}(x_{\n} J_{\m} +x_{\m} J_{\n} ) \Big) + \ldots \nn
\end{align}
which matches \eqref{AOP} with $b=0$, $e = -3/2$, determining the beta function for the background metric field according to \eqref{betafn}.

Computing the JJ OPE we find
\begin{align}
J_{\m}(x) J_{\n}(0)&\\ \sim \ldots +&\frac{1}{x^3} I_{\m\a}(x) \Big( (4 h^1_{\g\delta}(x) g_{\a\n} + 2 \tilde{h}_{\g\delta \a\n}(x))T^{\psi}_{\g\delta}(0) - 4 h^1_{\g\delta}(x) g_{\a\n} T^{\phi}_{\g\delta}(0) \Big) + \ldots \nn\\
 \sim \ldots +& \frac{1}{x^3}I_{\m\a}(x) \Big( (12 h^1_{\g\delta}(x) g_{\a\n} + 4 \tilde{h}_{\g\delta \a\n}(x)) (T^{\psi}_{\g\delta}(0)+T^{\phi}_{\g\delta}(0)) \nn\\+& 6( 4h^1_{\g\delta}(x) g_{\a\n} + \tilde{h}_{\g\delta \a\n}(x)) (\frac{1}{C_T^{\psi}}T^{\psi}_{\g\delta}(0)-\frac{1}{C_T^{\phi}}T^{\phi}_{\g\delta}(0)) \Big) + \ldots, \nn
\end{align}
where we again omitted descendants. The first term on the right-hand side is precisely $\frac{C_J}{C_T} I_{\m\a}(x) A_{\g\delta\a\n}(x) T_{\g\delta}(0)$ while the remainder gives a vanishing contribution to the $3$-point function $\left\langle T_{\m\n}(x_1) J_{\s}(x_2) J_{\r} (x_3)\right\rangle$. As we might have anticipated, a generic linear combination of $T_{\m\n}^{\phi}$ and $T_{\m\n}^{\psi}$ can appear in this OPE. However this can always be rewritten in terms of the true stress-energy tensor (i.e. the one obtained by varying the action with respect to the metric) plus another linear combination which is orthogonal to the stress-energy tensor. The beta function for the metric arises from the factor multiplying the true stress-energy tensor.

\subsection{Summary}

Let us conclude this section by summarising the general structure of the deformed theory. We observed that the singular terms in the JJ operator product expansion are associated with the renormalisation of the background metric and the emergence of Lifshitz symmetry. Using conformal perturbation theory,
the universal terms in this OPE give rise to a volume divergence and a divergence involving the stress energy tensor. Non-universal terms in the OPE can generate beta functions for other background fields (apart from the metric) which in general break the Lifshitz symmetry. Such additional terms in the OPE also imply that one cannot truncate to just the stress energy tensor and the vector operator, which will be reflected by the absence of a corresponding consistent truncation in the bulk.

In the two dimensional example, there were no non-universal divergences occurring in the OPE and therefore this case exactly realised the bulk scenario. In a typical higher dimensional model one may well obtain additional divergences and therefore running of associated background fields. In the first of our three dimensional examples there would be divergences arising from relevant operators such as $\phi^2$. We observed using our other 3d example that generically additional operators of dimension $d$ can also arise in the OPE, both descendants of lower dimension operators and primary operators which are orthogonal to the stress-energy tensor, and in this example there was a beta function for a second dimension $d$ operator.

\section{Conclusions} \label{sec:con}
In this paper we have developed holography for Lifshitz spacetimes with dynamical exponent $z=1+\e^2$, working perturbatively in $\epsilon$. We showed that the bulk theory is dual to a $d$-dimensional CFT deformed by a vector operator  of dimension $d$. Such a continuous deformation changes the relativistic fixed point into a non-relativistic one.

Conformal perturbation theory was used to study such deformations of a generic conformal field theory from the field theory perspective. We argued and demonstrated in specific examples that the Lifshitz invariance indeed appears generically in a deformed CFT. Without reference to any holographic dual, we could see directly from the vector operator OPEs that a renormalization of the background metric is induced by the vector deformation; this renormalization is responsible for the emergence of Lifshitz symmetry.

In standard QFT discussions, after deforming the CFT infinities give rise to beta functions and these drive the theory towards a new fixed point in the IR
where the beta functions vanish. In our discussion, the effect of the beta function is to change the condition of scale invariance from that of
relativistic invariance (i.e. tracelessness of the stress energy tensor) to Lifshitz invariance (i.e. vanishing of the $z$ weighted trace). In other words, as soon as the operator is switched on the theory finds itself in the non-relativistic fixed point rather than flowing to it in the IR.

Various extensions of this work would be possible. Firstly, deformations by other dimension $d$ tensor operators are likely to lead to similar results
since their OPEs should have a similar structure to that of the JJ OPE discussed here. In particular, one expects a logarithmic divergence proportional to the stress energy tensor, which should then induce a beta function for the metric. It would be interesting to systematically investigate all such possibilities.

The finite temperature behaviour of the system studied here will be explored in our subsequent paper \cite{Korovin2}. As mentioned at the beginning, one could also explore Lifshitz solutions with running scalar couplings (hyperscaling violation). In such a case the dual field theory may admit a similar interpretation as a specific deformation of a relativistic theory which either already has or acquires  a running coupling (corresponding to the hyperscaling violation).

Top down embeddings of Lifshitz solutions with $z \ge 1$ were found in \cite{Gregory:2010gx}\footnote{Note that there are typos in the uplift of the six-dimensional solutions to ten dimensions, which were pointed out in \cite{Jeong:2013jfc}. Alternative uplifts of these six-dimensional solutions to type IIB supergravity were found in \cite{Jeong:2013jfc}.}, with flows between such solutions being discussed in \cite{Braviner:2011kz}. Solutions with $z$ close to one are interpretable in terms of a vector deformation of the dual CFT \cite{Korovin2}. However, these solutions are obtained from
hyperboloidal reductions of (massive) gauged supergravities, which in turn can be uplifted to ten or eleven dimensions. In these cases the CFTs dual to the AdS solutions are not well understood; the reduction on the hyperboloid restricts the dynamical exponent $z$ to be discrete and there are clearly many additional fields in the bulk description relative to the Einstein-Proca model explored here. As discussed in the previous section, these fields (and their corresponding dual operators) are expected to be associated with deformations of the Lifshitz points. Despite these complications it is clearly interesting to explore the dual theories in these models using the techniques of this paper and these solutions  will be further discussed in \cite{Korovin2}.

Given that the bulk theory is relativistic, it is perhaps not surprising that the non-relativistic dual theory could be related to a relativistic field theory, but it is nonetheless highly non-trivial that they are continuously connected.
It has been argued in various works, see the recent discussions in
\cite{Janiszewski:2012nb,Griffin:2012qx}, that holography for Lifshitz theories would more naturally be described using non-relativistic Ho\v{r}ava-Lifshitz type gravities. It would be interesting to explore the relationship between the non-relativistic and relativistic descriptions for $z$ close to one. In particular, one might wonder whether they are dual to different universality classes of Lifshitz invariant theories.

\section*{Acknowledgments}

This work is part of the research program of the Stichting voor Fundamenteel Onderzoek der Materie (FOM), which is financially supported by the Nederlandse Organisatie voor Wetenschappelijk Onderzoek (NWO). YK and KS acknowledge support via an NWO Vici grant. KS and MT acknowledge support from a grant of the John Templeton Foundation. The opinions expressed in this publication are those of the authors and do not necessarily reflect the views of the John Templeton Foundation. We would like to thank Nikolay Bobev and Balt van Rees for collaboration during an early phase of this work.

\addtocontents{toc}{\protect\vspace*{\baselineskip}}



\appendix
\section{Useful formulae \label{ap1} }
Under a general transformation $g_{ij} \rightarrow g_{ij} + \delta g_{ij}$ the Christoffel symbols and Ricci tensors transform as:
\begin{align}
 \delta \G^{i}_{jk} &= \frac{1}{2} g^{il} (\nabla_j \delta g_{kl} + \nabla_k \delta g_{jl}-\nabla_l \delta g_{jk}), \label{dG} \\
 \delta R_{ij} &= \frac{1}{2} \Big(\nabla^k \nabla_i \delta g_{jk} + \nabla^k \nabla_j \delta g_{ik} - \nabla^2 \delta g_{ij} - \nabla_i \nabla_j \tr(\delta g_{ij}) \Big), \label{dRic}\\
\delta R &= \delta g^{ij} R_{ij} + g^{ij} \delta R_{ij}.
\end{align}
In particular, under a Weyl rescaling $\delta g_{ij} = 2 \sigma g_{ij}$ we get:
\begin{align}
 \delta \G^{i}_{jk} &=\delta^i_k \nabla_j \s + \delta^i_j \nabla_k \s - g_{jk} \nabla^i \s, \\
 \delta R_{ij} &= -g_{ij} \nabla^2 \s + (2-d) \nabla_i \nabla_j \s,\\
\delta R &= - 2 \s R + 2(1-d) \nabla^2 \s.
\end{align}

\section{Expansion of Einstein equations in d=2} \label{backr2d}

To obtain the coefficients in the expansion \eqref{expg[2]} we need to work out \eqref{Einst} at
order $\epsilon^2$.
Note that although $\epsilon$ appears in $M^2$ it always
multiplies a vector field and hence it can contribute only at higher order in the $\ep$ perturbation theory. Therefore to this order
we can set $M^2 =1$.
The rhs of \eqref{Einst} can be expanded near the boundary as follows
\begin{align}
e^{2r} t_{[2](0)ij} + e^{0r} (t_{[2](2)ij} + r \t_{[2](2)ij}) + \mathcal{O}(e^{-2r}),
\end{align}
where
\begin{align}
t_{[2](0)ij} &= A_{(0)i} A_{(0)j} - \frac{1}{2} A_{(0)k}A_{(0)}^k  g_{[0](0)ij}, \\
t_{[2](2)ij} &
= \frac{1}{2} \Big(A_{(0)i} a_{(2)j} + A_{(0)i} \nabla_j (\nabla_k A^k_{(0)})
+ (i\leftrightarrow j)\Big)  \\
&+\frac{1}{2} F_{(0)i}{}^{k} F_{(0)jk} - \frac{1}{4} F_{(0)kl}F^{kl}_{(0)}g_{[0](0)ij} \nn \\
&+ \Big(A_{(0)}{}^kA_{(2)k} - A_{(0)}{}^{k}a_{(2)k} - A_{(0)}{}^{k}\nabla_k (\nabla_n A^n_{(0)}) \Big)g_{[0](0)ij} \nn \\
&+
\frac{6 \pi}{c}(
\langle T^{kl} \rangle_{[0]} A_{(0)k}A_{(0)l} g_{[0](0)ij}
-
A_{(0)k}A^k_{(0)} \langle T_{ij} \rangle_{[0]}), \nn\\
\t_{[2](2)ij} & = A_{(0)}{}^{k}a_{(2)k} g_{[0](0)ij}.
\end{align}
Note also the identity
\begin{align}
\tr(t_{[2](2)ij} + r \t_{[2](2)ij}) = &2 A_{(0)i} A^i_{(2)} + (2 r-1) A_{(0)i}  a^i_{(2)} - A^i_{(0)} \nabla_i (\nabla_j A^j_{(0)}) \nn \\+&
\frac{12 \pi}{c} A^i_{(0)}\langle T_{ij}\rangle_{[0]} A^j_{(0)} - \frac{1}{4} R A^i_{(0)}A_{(0)i}.
\end{align}

The leading term in $r$ on the right-hand side of \eqref{Einst}
indicates that a logarithmic correction  $h_{[2](0)ij} r$
must be included at leading
radial order in $g_{[2]}$. It is given by \eqref{h20}.

In $d=2$ the following identities hold
\begin{align}
&\tr(h_{[2](0)})=0 \\
&\tr(g_{[0](2)}h_{[2](0)})=\frac{12 \pi}{c}
\tr(\langle T \rangle_{[0]}h_{[2](0)})
=\frac{1}{4} R A^i_{(0)}A_{(0)i}
- \frac{12 \pi}{c} A^i_{(0)} \langle T_{ij} \rangle_{[0]} A^j_{(0)}. \nn
\end{align}

Equation \eqref{Einstrr} allows us to compute $\tr(h_{[2](2)})$:
\begin{align}
\tr(h_{[2](2)}) &= \frac{1}{2} (\nabla_i A^i_{(0)})^2
- \frac{1}{4} F_{(0)}{}^{ij} F_{(0)ij}.
\end{align}
Note also the following useful relation:
\begin{align}
\tr(h_{[2](2)} - h_{[2](0)} g_{[0](2)}) = A_{(0)i} a^i_{(2)} + \frac{1}{2} \nabla_j \Big( A^j_{(0)}(\nabla_i A^i_{(0)}) -A_{(0)i} F^{ji}_{(0)} \Big).
\label{rel}
\end{align}
Collecting terms of order $\epsilon^2$ and $e^{0r}$ in \eqref{Einst} we get (for general $d$)
\begin{align}
\label{eqh}
&Ric_{[2]ij} \Big|_{r}  -\Big[(2-d)g_{[2](2)} + (2-d)r h_{[2](2)} +\frac{d-4}{2} h_{[2](2)}  \\
& \qquad + g_{[0](2)}g^{-1}_{[0](0)}h_{[2](0)} + h_{[2](0)}g^{-1}_{[0](0)}g_{[0](2)} \nn \\
& \qquad +\frac{1}{2} \tr \Big((2r-1) h_{[2](0)} g_{[0](2)} + (1-2r)h_{[2](2)} - 2g_{[2](2)} \Big) g_{[0](0)} \nn \\
& \qquad + \frac{R}{4} (1+2r) h_{[2](0)} \Big]_{ij} =t_{[2](2)ij} + r \t_{[2](2)ij}, \nn
\end{align}
where the trace is taken with $g^{-1}_{[0](0)}$. Taking
the trace of the last equation and using \eqref{dRic} and properties of Ricci tensor in $2$ dimensions
\begin{align}
 &Ric_{[2]ij}\Big|_{r}=\frac{r}{2} \Big(\nabla^k \nabla_j h_{[2](0)ik}+ \nabla^k \nabla_i h_{[2](0)jk} - \nabla^2h_{[2](0)ij} \Big)
\end{align}
we see that terms proportional to $r$ cancel separately and we obtain \eqref{trg2}.

Now we can solve for $h_{[2](2)ij}$ from \eqref{eqh}. The result is given in \eqref{h22}. The divergence of $g_{[2](2)ij}$ is determined
from the $e^{-2r}$ terms in \eqref{Einstri}, leading to
\begin{align}
\label{diver}
 &\nabla^j (h_{[2](2)ij} - 2 g_{[2](2)ij}) - \nabla_i \tr(h_{[2](2)} -2 g_{[2](2)} - g_{[0](2)} h_{[2](0)}) \\-&  g_{[0](2)}^{jk} \nabla_k h_{[2](0)ij}+\frac{6 \pi}{c} A^l_{(0)}A^k_{(0)} \nabla_i \langle T_{kl} \rangle_{[0]} - \frac{1}{4} A_{(0)i} A^j_{(0)} \nabla_j R  \nn\\=& - g_{[0](2)}^{jk} A_{(0)k} F_{(0)ij} + A_{(0)}^j (\partial_i A_{(2)j} - \partial_j A_{(2)i}) - A_{(2)i}(\nabla_j A^j_{(0)})\nn\\&+ \Big(a_{(2)}^j - A_{(2)}^j + \nabla^j(\nabla_k A^k_{(0)})\Big) F_{(0)ij}  + A_{(0)i} A_{(2)r}. \nn
\end{align}

\section{Expansion of Einstein equations in d=3} \label{Eq-analysis}

Collecting $\e^2$ terms (up to $e^{-r}$) in \eqref{Einst} gives (we keep explicit $d$ in our formulas, since some of our results apply to arbitrary dimension)
\begin{align}
\label{Rij}
&\quad e^{2r}\Big[-\frac{d}{2} h_{[2](0)} - \frac{1}{2}\tr(h_{[2](0)}) g_{[0](0)}\Big]_{ij}+ Ric_{[2]ij} \Big|_r \\
& + r \Big[(d-2) h_{[2](2)} + \tr(h_{[2](2)} - h_{[2](0)}g_{[0](2)}) g_{[0](0)} + \tr(g_{[0](2)})h_{[2](0)} \Big]_{ij} \nn\\
&+e^{0r} \Big[\frac{4-d}{2}h_{[2](2)} + (d-2)g_{[2](2)} -h_{[2](0)} g_{[0](0)}^{-1} g_{[0](2)} - g_{[0](2)} g_{[0](0)}^{-1} h_{[2](0)}  \nn\\
& \qquad +\frac{1}{2}\tr(2g_{[2](2)} - h_{[2](2)} + h_{[2](0)} g_{[0](2)})g_{[0](0)} +\frac{1}{2}\tr(g_{[0](2)})h_{[2](0)} \Big]_{ij} \nn\\
&+ r e^{-r} \Big[\frac{3 (d-3)}{2}h_{[2](3)} +\frac{3}{2} \tr(h_{[2](3)} -h_{[2](0)}g_{[0](3)})g_{[0](0)} \Big]_{ij} \nn\\
& + e^{-r} \Big[\frac{6-d}{2}h_{[2](3)} + \frac{3 (d-3)}{2}g_{[2](3)} - \frac{3}{2}\Big( h_{[2](0)} g_{[0](0)}^{-1} g_{[0](3)} + g_{[0](3)} g_{[0](0)}^{-1} h_{[2](0)} \Big)\nn\\
& \qquad +\frac{1}{4}\tr(h_{[2](0)})g_{[0](3)} +\frac{1}{2} \tr(3g_{[2](3)} +h_{[2](0)}g_{[0](3)} -h_{[2](3)})g_{[0](0)} \Big]_{ij} \nn\\
&=e^{2r} \Big[ \frac{d}{2} A_{(0)i} A_{(0)j} + \frac{1}{2 (1-d)} A_{(0)k} A^k_{(0)} g_{[0](0)ij} \Big]+e^{0r} \Big[ \frac{1}{2} g^{kl}_{[0](0)}F_{(0)ik} F_{(0)jl} \nn\\
&+\frac{d-2}{2} (A_{(0)i} A_{(2)j} + A_{(0)j} A_{(2)i}) -\frac{1}{2}(A_{(0)i} \nabla_j A_{(0)r} + A_{(0)j} \nabla_i A_{(0)r}) \nn\\
&+\frac{1}{4(1-d)} \Big( -4(A_{(0)k} A^k_{(2)} + A_{(0)k} \nabla^k A_{(0)r})g_{[0](0)ij} + F_{(0)kl} F^{kl}_{(0)} g_{[0](0)ij} \nn\\
&-2A^k_{(0)} g_{[0](2)kl} A^l_{(0)}g_{[0](0)ij} + 2A_{(0)k} A^k_{(0)} g_{[0](2)ij} \Big)  \Big]+r e^{-r} a_{(3)k} A_{(0)}^k g_{[0](0)ij}  \nn\\
&+e^{-r} \Big[ \frac{d-3}{2} (A_{(0)i} A_{(3)j} + A_{(0)j} A_{(3)i}) + \frac{2}{d-1}A_{(0)k} A^k_{(3)}g_{[0](0)ij} \nn\\
&\qquad+\frac{1}{2(d-1)} A^k_{(0)} g_{[0](3)kl} A^l_{(0)}g_{[0](0)ij} -\frac{1}{2(d-1)} A_{(0)k} A^k_{(0)} g_{[0](3)ij} \nn\\
&\qquad+\frac{1}{2}\Big( a_{(3)i} A_{(0)j} + a_{(3)j} A_{(0)i} - a_{(3)k} A_{(0)}^k g_{[0](0)ij} \Big) \Big]. \nn
\end{align}
The trace of it in $d=3$ gives:
\begin{align}
\label{trRij}
 &\tr(Ric_{[2]}|_r) - e^{2r} \Big[3 \tr(h_{[2](0)}) + (-4r+1)e^{-2r} \tr(h_{[2](2)})  \\&- 4 e^{-2r} \tr(g_{[2](2)})+(3r+\frac{1}{2}) e^{-2r} \tr(g_{[0](2)} h_{[2](0)})-(r+\frac{1}{2}) e^{-2r} \tr(g_{[0](2)}) \tr(h_{[2](0)})\nn\\&- \frac{9}{2}r e^{-3r} \tr(h_{[2](3)}) - \frac{9}{2} e^{-3r} \tr(g_{[2](3)}) +\frac{1}{2}(9r+3) e^{-3r} \tr(g_{[0](3)} h_{[2](0)})\Big]\nn\\
&=\frac{3}{4} e^{2r} A_{(0)i} A^i_{(0)} + e^{0r} \Big(\frac{5}{2} A_{(0)i} A^i_{(2)} -\frac{1}{4}A^i_{(0)} \nabla_i \nabla_j A^j_{(0)} +\frac{1}{8}F_{(0)ij} F^{ij}_{(0)} \nn\\&+ \frac{3}{4}A^i_{(0)} g_{[0](2)ij} A^j_{(0)} - \frac{1}{4} \tr(g_{[0](2)})A_{(0)i} A^i_{(0)}  \Big) + 3 r e^{-r} a_{(3)i} A_{(0)}^i  \nn\\&+ e^{-r}\Big(3 A_{(0)i} A^i_{(3)} + \frac{3}{4} A^i_{(0)} g_{[0](3)ij} A^j_{(0)} - \frac{1}{2} a_{(3)i} A_{(0)}^i \Big). \nn
\end{align}
The $R_{rr}$ equation \eqref{Einstrr} gives:
\begin{align}
\label{Rrr}
 &-\tr(h_{[2](0)}) + e^{-2r}\tr(h_{[2](2)}) + \frac{3}{2} r e^{-3r} \tr(g_{[0](3)}h_{[2](0)} - h_{[2](3)})\\&+ e^{-3r} \tr(2 h_{[2](3)} -\frac{1}{2}g_{[0](3)}h_{[2](0)}- \frac{3}{2} g_{[2](3)})\nn\\
&= \frac{1}{4} A_{(0)i} A_{(0)}^i \nn\\
&+\frac{1}{4} e^{-2r}\Big((\nabla_i A_{(0)}^i)^2 - A_{(0)}^i g_{[0](2)ij} A_{(0)}^j - 2 A_{(0)i}A_{(2)}^i  + A_{(0)}^i \nabla_i \nabla_j A_{(0)}^j - \frac{1}{2} F_{(0)ij} F_{(0)}^{ij}\Big) \nn\\
&-r e^{-3r} a_{(3)i} A_{(0)}^i +e^{-3r} \Big( -\frac{1}{4}A_{(0)}^i g_{[0](3)ij} A_{(0)}^j -A_{(0)i}A_{(3)}^i  + \frac{1}{2} a_{(3)i} A_{(0)}^i \Big). \nn
\end{align}
The $R_{ri}$ Einstein equation \eqref{Einstri} reads at order $\e^2$
\begin{align}
\label{3Rri}
&\nabla^j h_{[2](0)ij} - \nabla_i \tr(h_{[2](0)}) \\
+ &r e^{-2r} \Big(  2 h_{[2](0)}^{jk} \nabla_k g_{[0](2)ij} + g_{[0](2)}^{jk} \nabla_i h_{[2](0)jk} - 2 \nabla^j h_{[2](2)ij} \nn\\&+ g_{[0](2)ik} (2 \nabla_j h_{[2](0)}^{jk} - \nabla^k \tr(h_{[2](0)})) - 2 \nabla_i \tr(g_{[0](2)} h_{[2](0)} - h_{[2](2)})   \Big) \nn\\
+& e^{-2r} \Big( \nabla^j h_{[2](2)ij} - 2 \nabla^j g_{[2](2)ij} -g_{[0](2)}^{jk} \nabla_k h_{[2](0)ij} - \frac{1}{2}h_{[2](0)}^{jk} \nabla_i g_{[0](2)jk} \nn\\&-\frac{1}{2} h_{[2](0)ik} (2 \nabla_j g_{[0](2)}^{jk} - \nabla^k \tr(g_{[0](2)})) - \nabla_i \tr(h_{[2](2)} - 2 g_{[2](2)} - g_{[0](2)} h_{[2](0)})  \Big) \nn\\
+& r e^{-3r} \Big( 3 h_{[2](0)}^{jk} \nabla_k g_{[0](3)ij} + \frac{3}{2} g_{[0](3)}^{jk} \nabla_i h_{[2](0)jk} - 3 \nabla^j h_{[2](3)ij} \nn\\&+ \frac{3}{2} g_{[0](3)ik} (2 \nabla_j h_{[2](0)}^{jk} - \nabla^k \tr(h_{[2](0)})) - 3 \nabla_i \tr(g_{[0](3)} h_{[2](0)} - h_{[2](3)}) \Big) \nn\\
+& e^{-3r} \Big( \nabla^j h_{[2](3)ij} - 3 \nabla^j g_{[2](3)ij} -g_{[0](3)}^{jk} \nabla_k h_{[2](0)ij} - \frac{1}{2}h_{[2](0)}^{jk} \nabla_i g_{[0](3)jk} \nn\\&- \nabla_i \tr(h_{[2](3)} - 3 g_{[2](3)} - g_{[0](3)} h_{[2](0)})  \Big) \nn\\
&= 2 A_{(0)i} A_{(0)r} \nn\\
&+e^{-2r} \Big( 2(A_{(2)i} A_{(0)r} + A_{(0)i} A_{(2)r}) - g_{[0](2)}^{jk} F_{(0)ij} A_{(0)k} \nn\\&+ A_{(0)}^j (\nabla_i A_{(2)j} - \nabla_j A_{(2)i}) - A_{(2)}^j F_{(0)ij} + \frac{1}{2} F_{(0)ij} \nabla^j \nabla_k A_{(0)}^k   \Big) \nn\\
&+ r e^{-3r} \Big( 2( a_{(3)i} A_{(0)r} + a_{(3)r} A_{(0)i}) +  A_{(0)}^j (\nabla_i a_{(3)j} - \nabla_j a_{(3)i}) - 2 a^j_{(3)} F_{(0)ij} \Big) \nn\\
&+ e^{-3r} \Big( 2 (A_{(3)i} A_{(0)r} + A_{(3)r} A_{(0)i}) -g_{[0](3)}^{jk} F_{(0)ij} A_{(0)k} \nn\\&+  A_{(0)}^j (\nabla_i A_{(3)j} - \nabla_j A_{(3)i}) - 2 A^j_{(3)} F_{(0)ij} +  a^j_{(3)} F_{(0)ij}\Big). \nn
\end{align}

\section{Normalization of the current \label{ap2}}

To fix the normalization of the current in the QFT dual to the Lifshitz geometry in three bulk dimensions we need to compute the $2$-point function $\left\langle J_i(x) J_j(0) \right\rangle $ holographically and compare its normalization to that in the CFT. Our discussion follows \cite{Mueck:1998iz}: note that in \cite{Mueck:1998iz} only the case of non-integer conformal dimension was considered but the result can be extracted from there by analytic continuation.

The Euclidean action for the bulk vector field is 
\be
S = \frac{1}{16 \p G_{d+1}} \int d^{d+1}x \sqrt{G} \left [ \frac{1}{4} F^{2} + \frac{1}{2} m^2 A^2 \right ].
\ee
Note that the action in Euclidean signature acquires an overall minus sign. 
With such a normalization the two point function in $d=2$ for the dual operator of dimension two is  \cite{Mueck:1998iz}
\begin{align}
\left\langle J_i(\vec{x}) J_j(0) \right\rangle = \frac{1}{4 \p G_3} \frac{1}{\p x^4}\Big( \delta_{ij} - 2 \frac{x_i x_j}{x^2}\Big).
\label{J2ptfn}
\end{align}
In complex coordinates $z= x_1 + i x_2$, $\bar{z} = x_1 - i x_2$ the form $\Big( \delta_{ij} - 2 \frac{x_i x_j}{x^2}\Big)$ equals
\[ \left( \begin{array}{cc}
-\frac{z^2 + \bar{z}^2}{2 z \bar{z}} & i\frac{z^2 - \bar{z}^2}{2 z \bar{z}}  \\
i\frac{z^2 - \bar{z}^2}{2 z \bar{z}} & \frac{z^2 + \bar{z}^2}{2 z \bar{z}}  \end{array} \right).\]
Using $J_z = \frac{1}{2}(J_1-iJ_2)$ we get
\begin{align}
 \left\langle J_z J_z\right\rangle = \frac{1}{4} (\left\langle J_1 J_1\right\rangle - 2 i \left\langle J_1 J_2\right\rangle - \left\langle J_2 J_2\right\rangle).
\end{align}
If the two point function $\left\langle J_i J_j\right\rangle = \frac{C_J}{x^4}\Big( \delta_{ij} - 2 \frac{x_i x_j}{x^2}\Big)$ one obtains
\begin{align}
 \left\langle J_z J_z\right\rangle = - \frac{C_J}{2}\frac{1}{z^3 \bar{z}},
\end{align}
which has a sign in agreement with our CFT computation in \eqref{genfunct}.
Comparing \eqref{J2ptfn} to the CFT normalization we can then fix $C_J = \frac{1}{4 \p^2 G_3}$.

\section{Stress-energy tensor correlation functions in two-dimensional Lifshitz theory.} \label{apcorr}

In this appendix we give the $2$-point correlation functions of the conserved stress-energy tensor in the two dimensional theory. Starting from $\left\langle J_{tt}(t,x) J_{tt}(0)\right\rangle $, $\left\langle J_{xt}(t,x) J_{tt}(0)\right\rangle $ (which are given in \eqref{ttttcorr}, \eqref{xtttcorr}) and $\left\langle J_{xx}(t,x) J_{xx}(0)\right\rangle $ (see below)  and applying the diffeomorphism Ward identity one can derive all other correlation functions. In terms of the quantity $f_{\m\n,\r\s}$ defined in \eqref{fcorr} the relevant identities are
\begin{align}
f'_{tx,tt}(\c) &= 4zf_{tx,tx}(\c) + z \c f'_{tx,tx}(\c), \\
f'_{xt,xx}(\c) &= 2(1+z)f_{xx,xx}(\c) + z \c f'_{xx,xx}(\c), \\
f'_{xt,xt}(\c) &= (3+z)f_{xt,xx}(\c) + z \c f'_{xt,xx}(\c), \\
f'_{xt,tt}(\c) &= 2(1+z)f_{xt,tx}(\c) + z \c f'_{xt,tx}(\c), \\
f'_{xt,tt}(\c) &= 2(1+z)f_{xx,tt}(\c) + z \c f'_{xx,tt}(\c), \\
f'_{xt,tx}(\c) &= (1+3z)f_{xx,tx}(\c) + z \c f'_{xx,tx}(\c).
\end{align}
These relations together with the symmetry of $f$ with respect to the exchange of the first pair of indices with the second pair provide enough information to construct all $2$-point correlation functions. We summarize the results here:
\begin{align}
&\left\langle J_{xx}(t,x) J_{xx}(0)\right\rangle = \frac{c}{(2 \p)^2}  x^{-2(1+z)} \Big[\frac{ \c^4 - 6\c^2 +1 }{(\c^2 +1)^4} \\&-\frac{2\e^2}{3} \frac{2\c^6 - 43 \c^4 + 58 \c^2 -5}{(\c^2+1)^5}
+\e^2 \frac{(\c^2-1) (\c^4 - 14 \c^2 +1)}{(\c^2+1)^5} \log(1 + \c^2)  \Big],\nn \\
&\left\langle J_{xt}(t,x) J_{xx}(0)\right\rangle = -4\frac{c}{(2 \p)^2}  x^{-2-(1+z)} \Big[\frac{\c(\c^2 -1)}{(\c^2 +1)^4} \\&-\frac{\e^2}{6} \frac{\c(2\c^4 - 5 \c^2 +1}{(\c^2+1)^5}
+\e^2 \frac{\c (2\c^4 -5 \c^2 +1)}{(\c^2+1)^5} \log(1 + \c^2)  \Big],\nn \\
&\left\langle J_{tx}(t,x) J_{xx}(0)\right\rangle = -4\frac{c}{(2 \p)^2}  x^{-(1+z)-2z} \Big[\frac{\c(\c^2 -1)}{(\c^2 +1)^4} \\&-\frac{\e^2}{6} \frac{\c(2\c^4 - 5 \c^2 +1)}{(\c^2+1)^5}
+\e^2 \frac{\c (\c^4 -5 \c^2 +2)}{(\c^2+1)^5} \log(1 + \c^2) \Big],\nn\\
&\left\langle J_{tx}(t,x) J_{tx}(0)\right\rangle = \frac{c}{(2 \p)^2}  x^{-4z} \Big[-\frac{ \c^4 - 6\c^2 +1 }{(\c^2 +1)^4}\\&+\frac{\e^2}{6} \frac{9 \c^6 - 209\c^4 + 203\c^2 -11}{(\c^2+1)^5}
+2\e^2 \frac{5 \c^4 - 10 \c^2+1}{(\c^2+1)^5} \log(1 + \c^2)  \Big],\nn
\end{align} 
\begin{align}
&\left\langle J_{xt}(t,x) J_{tx}(0)\right\rangle = \frac{c}{(2 \p)^2}  x^{-2-2z} \Big[-\frac{ \c^4 - 6\c^2 +1 }{(\c^2 +1)^4}\\&+\frac{\e^2}{3} \frac{(\c^2-1)(7\c^4 -94\c^2 +7)}{(\c^2+1)^5}
-\e^2 \frac{(\c^2-1)(\c^4 - 14 \c^2+1)}{(\c^2+1)^5} \log(1 + \c^2)  \Big],\nn\\
&\left\langle J_{xt}(t,x) J_{xt}(0)\right\rangle = \frac{c}{(2 \p)^2}  x^{-4} \Big[-\frac{ \c^4 - 6\c^2 +1 }{(\c^2 +1)^4}\\&+\frac{\e^2}{6} \frac{11\c^6 -203\c^4 +209\c^2 -9}{(\c^2+1)^5}
-2\e^2 \frac{\c^2(\c^4 - 10 \c^2+5)}{(\c^2+1)^5} \log(1 + \c^2)  \Big],\nn\\
&\left\langle J_{tt}(t,x) J_{xx}(0)\right\rangle = \frac{c}{(2 \p)^2}  x^{-2(1+z)} \Big[-\frac{ \c^4 - 6\c^2 +1 }{(\c^2 +1)^4}\\&+\frac{\e^2}{3} \frac{(\c^2-1)(7\c^4 -94\c^2 +7)}{(\c^2+1)^5}
-\e^2 \frac{(\c^2-1)(\c^4 - 14 \c^2+1)}{(\c^2+1)^5} \log(1 + \c^2)  \Big].\nn
\end{align}

\section{Scheme-dependence and Weyl invariance  \label{ap3}}
Here we collect some formulas which ae used in our discussion of scheme dependence and the check of Weyl invariance. We will consider three specific integrals: $\int \sqrt{\g} F_{ij} F^{ij}$, $\int \sqrt{\g} (\nabla_i A^i)^2$, $\int \sqrt{\g} R A^2$. These are finite in two dimensions and thus one could have added these as finite counterterms. In this section we consider contributions of these terms to one-point functions, their Weyl transformations and their possible effect on Ward identities.

Let us begin with $\int \sqrt{\g} F_{ij} F^{ij}$. The variation of such an action is
\begin{align}
 \delta \int \sqrt{\g} F_{ij} F^{ij} = \int \sqrt{\g} \Big[ -4 \nabla_i F^{ij} \delta A_j + (2 F_{ik} F_{jl} \g^{kl} - \frac{1}{2} F_{kl} F^{kl} \g_{ij})\delta \g^{ij} \Big].
\end{align}
We hence obtain the following contributions to one-point functions:
\begin{align}
\langle J^i \rangle &= -4 \nabla_k F_{(0)}^{ki}, \\
\langle T_{ij} \rangle &= 4 F_{(0)ik} F_{(0)jl} g_{[0](0)}^{kl} - F_{(0)kl} F_{(0)}^{kl} g_{[0](0)ij}.
\end{align}
The Weyl variation is
\begin{align}
\delta_W [F_{ij} F^{ij}] = - 2\s F_{ij} F^{ij} + 4 F^{ij} A_j \nabla_i \s.
\end{align}
The trace of the stress-energy tensor is:
\begin{align}
\langle T_{i}^i \rangle &= 2 F_{(0)ij} F_{(0)}^{ij}.
\end{align}
Next we consider $\int \sqrt{\g} (\nabla_i A^i)^2$. Its variation is
\begin{align}
 \delta \int \sqrt{\g} (\nabla_i A^i)^2 &= \int \sqrt{\g} \Big[ -2 \nabla^j(\nabla_i A^i) \delta A_j \\+ \Big(\frac{1}{2}& (\nabla_k A^k)^2 \g_{ij} + A^k \nabla_k \nabla_l A^l \g_{ij} - 2 A_j \nabla_i \nabla_k A^k\Big) \delta \g^{ij} \Big]. \nn
\end{align}
From here we get the following contributions to one-point functions:
\begin{align}
\langle J^i \rangle &= -2 \nabla^j(\nabla_i A_{(0)}^i), \\
\langle T_{ij} \rangle &=  (\nabla_k A_{(0)}^k)^2 g_{[0](0)ij} + 2 A_{(0)}^k \nabla_k \nabla_l A_{(0)}^l g_{[0](0)ij} - 4 A_{(0)j} \nabla_i \nabla_k A_{(0)}^k.
\end{align}
The Weyl variation is
\begin{align}
\delta_W [(\nabla_i A^i)^2] = - 2\s (\nabla_i A^i)^2 + 2 (\nabla_j A^j) A_i \nabla_i \s.
\end{align}
The trace of the stress-energy tensor is:
\begin{align}
\langle T_{i}^i \rangle &= 2 (\nabla_i A^i)^2.
\end{align}
Finally, consider $\int \sqrt{\g} R A^2$. Its variation is
\begin{align}
 \delta \int \sqrt{\g} R A^2 &= \int \sqrt{\g} \Big[2R A^i \delta A_i+ R A_i A_j \delta \g^{ij} \Big].
\end{align}
From here we obtain contributions to one-point functions:
\begin{align}
\langle J^i \rangle &= 2R A^i, \\
\langle T_{ij} \rangle &=  2R A_i A_j.
\end{align}
The Weyl variation is
\begin{align}
\delta_W [R A^2] = - 2\s R A^2 - 2 A^2 \Box \s.
\end{align}
The trace of the stress-energy tensor is:
\begin{align}
\langle T_{i}^i \rangle &= 2 R A^2.
\end{align}
Putting these results together it is also straightforward to check the Weyl invariance of the $d=2$ analogue of the Deser-Nepomechie action:
\begin{align}
 L_{DN} = - \frac{12 \p}{c}A^i_{(0)} \langle T_{ij} \rangle_{[0]} A^j_{(0)} + \frac{1}{4} F_{(0)ij}F_{(0)}^{ij}-\frac{1}{2}(\nabla_i A^i_{(0)})^2 + \frac{R}{4}A^i_{(0)} A_{(0)i}.
\end{align}


\addtocontents{toc}{\protect\vspace*{\baselineskip}}

\providecommand{\href}[2]{#2}\begingroup\raggedright\endgroup


\end{document}